\documentclass[useAMS,usenatbib]{mnras}
\usepackage{newtxtext,newtxmath} 
\usepackage[T1]{fontenc}
\usepackage{ae,aecompl}
\usepackage{todonotes}
\usepackage{mathtools}

\usepackage{graphicx}	% Including figure files
\usepackage{amsmath}	% Advanced maths commands
\usepackage{amssymb}	% Extra maths symbols

\newcommand{\myr}{\mbox{${\rm Myr}$}}
\newcommand{\pc}{\mbox{${\rm pc}$}}
\newcommand{\msun}{\mbox{M$_\odot$}}
\newcommand{\dex}{{\rm dex}}
\newcommand*\mean[1]{\overline{#1}}

\title[Cluster disruption by tidal shocks]{A systematic analysis of star cluster disruption by tidal shocks -- I.~Controlled $N$-body simulations and a new theoretical model}

\author[Webb, Reina-Campos \& Kruijssen]{
Jeremy J. Webb,$^1$\thanks{E-mail: webb@astro.utoronto.ca (JJW); reina.campos@uni-heidelberg.de (MRC); kruijssen@uni-heidelberg.de (JMDK)}
Marta Reina-Campos,$^{2\star}$
and J. M. Diederik Kruijssen$^{2\star}$
\\
$^{1}$ Department of Astronomy and Astrophysics, University of Toronto, 50 St. George Street, Toronto, ON, M5S 3H4, Canada\\
$^{2}$ Astronomisches Rechen-Institut, Zentrum f\"{u}r Astronomie der Universit\"{a}t Heidelberg, M\"{o}nchhofstra\ss e 12-14, 69120 Heidelberg, Germany\\
}

\begin{document}
\date{}
\pagerange{\pageref{firstpage}--\pageref{lastpage}} \pubyear{2018}

\maketitle
\label{firstpage}

\begin{abstract}
Understanding the evolution of stellar clusters in an evolving tidal field is critical for studying the disruption of stellar clusters in a cosmological context. We systematically characterise the response of stellar clusters to tidal shocks using controlled $N$-body simulations of clusters with various properties that are subjected to different types of shocks. We find that the strength of the shock and the density of the cluster within the half-mass radius are the dominant properties that drive the amount of mass lost by the cluster, with the shape of the cluster profile being of minor influence. When the shock is applied as two separate sub-shocks, the amount of mass loss during the second sub-shock is sensitive to the gap time between them. Clusters that experience successive sub-shocks separated by less than their crossing time attain the same masses and sizes at the end of the simulation. However, clusters subjected to sub-shocks separated by more than a crossing time experience different evolutionary histories. The amount of mass lost in the $N$-body models and its scaling with shock and cluster properties differs from that predicted by classical tidal disruption theory. We demonstrate that the discrepancy is alleviated by including a dependence on the escape time-scale of unbound stars, analogously to mass loss driven by two-body relaxation. With our new theoretical model for shock-driven mass loss, the predicted relative amounts of mass loss agree with the results of the $N$-body simulations to $\sim0.3~\dex$ across the full suite of simulations.

%Removed (50~per~cent)

\end{abstract}


\section{Introduction}

The long-term evolution of a star cluster is governed by both internal and external processes. External mechanisms, which depend on the cluster's formation environment and subsequent orbit within its host galaxy, include tidal stripping, tidal shocks, and dynamical friction. Tidal shocks are well-known to play an important role in the dynamical evolution of star clusters, both during formation and their long term evolution \citep[e.g.][]{spitzer58,spitzer87,ostriker72,chernoff86,aguilar88,chernoff90,kundic95,gieles06,kruijssen11}. In many (and possibly most) cases, tidal shock-driven mass loss dominates over the other mass loss mechanisms across a cluster's history \citep{lamers06,elmegreen10,elmegreen10b,kruijssen15b,miholics17,pfeffer18,li18}.

Over the course of a cluster's lifetime, shocks can occur due to interactions with any form of granularity in the gravitational potential, such as giant molecular clouds \citep[GMCs,][]{gieles06,lamers06}, spiral arm passages \citep{gieles07}, galaxy merger-induced structure \citep{kruijssen12}, passages through the Galactic disc \citep{gnedin97,kruijssen09,webb14b}, and perigalactic passes \citep{gnedin97,baumgardt03,webb13, webb14a}. A cluster orbiting in any realistic tidal field that contains substructure (as opposed to a smooth distribution of matter) will be subject to a non-negligible number of tidal shocks. Hence, obtaining a systematic understanding of how individual shocks influence cluster evolution is an important step towards understanding the distribution of cluster properties observed today throughout the Universe. Furthermore, being able to quantify the effects of tidal shocks will not only enable probing the conditions of the formation environment and birth properties of stellar clusters, but it will also allow for present day clusters to be used as tools to study the formation and assembly history of their host galaxy \citep[e.g.][]{kruijssen19b,kruijssen19c}.

The earliest framework for quantifying how clusters are affected by tidal shocks dates back to \citet{spitzer58}, who used the impulse approximation to explore how tidal interactions with GMCs determine a cluster's disruption time. \citet{spitzer58} found that repeated GMC interactions accelerate how quickly a cluster will dissolve, with the change in energy and the amount of mass lost during an interaction dependent on cluster density and the amount of energy injected by the encounter. \citet{aguilar85} performed a series of $N$-body simulations of collisionless systems to determine how well the impulse approximation predicted the changes in mass experienced by interacting spherical galaxies. The authors found that, when integrating over the entire orbital path of the perturbing galaxy, the impulse approximation could accurately predict the change in energy and mass of the primary galaxy. In fact, the mass and density evolution can be predicted given the precise details of each shocking event \citep{aguilar86}. The authors also explored the validity of using the tidal approximation, which assumes the impact parameter is sufficiently large such that the density profile of the perturber can be ignored, which allows to determine analytically the impulse given to each star. \citet{aguilar85} find that the change in energy of stars that remain bound after a shock and the overall change in mass could not be recovered when using the impulse and tidal approximations together. The fundamental issue with the tidal approximation is that it does not take into consideration the distribution of stellar velocities in the system being perturbed, which \citet{aguilar85} note is what prevents the existence of a clear scaling relation between changes in mass and the properties of the two galaxies. However, when the exact details of individual shocking events are unknown, the tidal approximation is unavoidable and it is often assumed that the relative change in mass is proportional to the relative change in energy.

\citet{gieles06} has since updated the impulse approximation, showing that the inclusion of gravitational focusing and accounting for close encounters further adds to a GMCs ability to remove stars from a cluster. The description of shock-driven mass loss was refined further by \citet{kruijssen11}, who included the second-order term in the shock-driven mass loss rate. Even more recently, \citet{gieles16} have included the cluster density evolution under the combined influence of repeated tidal shocks and two-body relaxation, enabling a better estimate of the disruption time for large statistical ensembles of shocks. Based on these works, it is possible to determine the time-scale over which a cluster is disrupted due to tidal shocks of any kind (e.g.~GMCs, disc crossings, spiral arm crossings) and model the mass evolution of individual clusters along a given orbit \citep[e.g.][]{ostriker72, gieles07, kruijssen09}. However, in each of the cases, it is necessary to assume that the local density, mass, structure, and relative velocity dispersion of the perturbing sources, as well as the fraction ($f$) of the injected energy that contributes to the cluster mass loss ($\Delta M/M = f \Delta E/E$), are all identical and static in order to analytically estimate the cluster disruption time.

A more flexible method for determining the time-scale over which tidal shocks drive cluster disruption stems from extracting the tidal tensor experienced by a cluster as a function of time from galaxy simulations \citep{prieto08,kruijssen11}. By integrating each element of the tidal tensor over the duration of a single shock, one can determine the integrated tidal heating experienced by the cluster, which defines its change in energy over the course of a tidal shock. By defining a `tidal heating parameter' \citep{gnedin03} and again assuming $\Delta M/M = f \Delta E/E$, a cluster's mass loss history can be estimated for any known tidal history. Such an approach is ideal for modelling a single shock event or multiple events where the tidal shock sources are neither static or uniform and the evolution of the tidal tensor can be determined. This approach is difficult to apply to clusters observed in the present-day Universe, as their complete tidal histories are typically unknown. However, in cases where the tidal tensor is known explicitly (such as in models of galaxy formation and evolution), the effects of all types of tidal shocks can be traced.

To date, no dynamically motivated framework for how individual shocks affect star clusters has been successfully tested against direct $N$-body simulations in cases where the tidal approximation is necessary. The need for such a model is of increasing importance as the resolution difference between large-scale cosmological simulations of galaxy formation and small-scale simulations of star cluster evolution is continually decreasing. In fact, several large-scale simulations of galaxy formation have reached the resolution scale necessary to identify sites of star cluster formation and track the location of star cluster particles as galaxies form and evolve \citep[e.g.][]{kravtsov05, maxwell12, li17, pfeffer18, kim18, mandelker18}. Since the tidal fields experienced by such clusters evolve with time and contain substructure in the form of GMCs, spiral arms, discs, and spheroids, the clusters are subject to a wide range of tidal shock strengths, frequencies, and durations over their lifetimes. Hence, a cluster's mass loss rate cannot be estimated by assuming a single set of global properties for the perturbing source. Furthermore, since individual clusters are not directly modelled in these large scale simulations, the full form of the impulse approximation cannot be applied either. A general framework for how clusters respond to tidal shocks is needed before the ever increasing resolution of cosmological simulations can be exploited.

Understanding how the evolution of a cluster in a cosmologically motivated tidal field compares to a cluster orbiting in a static and smooth tidal field is necessary before models of star clusters can be accurately compared to observations. An accelerated disruption time due to tidal shocks has profound implications for the evolution of the globular cluster\footnote{Note that we do not distinguish between open and globular clusters in this work. Dynamically, the tidal shock-driven mass loss rate depends on the cluster density rather than its mass. Open and globular clusters span a similar range in densities \citep[e.g.][]{krumholz19}, implying that it is not necessary to distinguish between these systems.} mass function \citep[GCMF][]{elmegreen10,kruijssen15b,reinacampos18}, the stellar mass function \citep[MF][]{vesperini97,baumgardt03,kruijssen09c, lamers13, webb14a, baumgardt17} of individual clusters, and the distribution of cluster sizes \citep{webb14a,webb14b,gieles16}. If tidal shocks have had a stronger effect on cluster evolution in the past, estimates of the initial GCMF based on the present day distribution of cluster masses and the tidal field of the galaxy underestimate the mass and number of clusters that form in high-redshift galaxies. Furthermore, if a proper treatment of tidal shocks (e.g.\ from interactions with GMCs in their natal environment) results in clusters experiencing higher mass loss rates, current estimates of initial cluster masses based on their orbit \citep{gnedin97,dinescu99,baumgardt03} are in fact lower limits. This second point may offer an explanation for why the stellar MFs of Galactic clusters suggest they have lost a higher fraction of their initial mass than their present-day orbit indicates \citep{webb15}. Finally, tidal shocks experienced by tidally under-filling clusters will result in cluster expansion as opposed to mass loss. Hence tidal shocks may help clusters expand from compact sizes at formation to their present day sizes, offering a potential pathway for producing extended (globular) clusters. 

In this work, we present a systematic study of how individual tidal shocks and successive pairs of shocks affect stellar clusters. For this purpose, we use controlled $N$-body simulations in which we subject clusters to a variety of tidal shocks. Our approach specifically focuses on quantifying the amount of mass lost by the cluster due to a tidal shock, i.e.~how changes in the tidal tensor along the cluster's orbit causes stars to become unbound. As previously discussed, the use of the tidal tensor offers a way of quantifying the strength of the background tidal field that does not require an analytic fit to the host galaxy's potential and it can evolve with time. A tidal tensor-based analysis is particularly useful for studies that model the evolution of stellar clusters in a cosmological context, during which interactions with GMCs dominate cluster mass loss, as an on-the-fly determination of the tidal tensor at the location of the star cluster particles allows for a more accurate description of their evolution \citep[e.g.][]{pfeffer18}.

The structure of this paper is as follows. In Section~\ref{sec:nbody}, we introduce the suite of simulations of star clusters that are subjected to a range of different tidal shocks. Section~\ref{sec:results} explores the mass and size evolution of each model, including a discussion of how the mass lost by each cluster during a shock depends on the detailed evolution of the cluster's properties and of the tidal tensor over the course of the shock. In Section~\ref{sec:discussion}, we present a new, dynamically-motivated theoretical model for cluster mass loss due to tidal shocks, which accurately reproduces the $N$-body simulations. Finally, we summarise and discuss our conclusions in Section~\ref{sec:conclusion}.

\section {$N$-body Simulations} \label{sec:nbody}

In order to explore the amount of mass lost by a stellar cluster after a tidal shock, and how this depends on the properties of the shock and the cluster itself, we perform a large number of direct $N$-body simulations with \texttt{NBODY6tt} \citep{aarseth03,aarseth10,renaud11} with different shock durations and strengths, as well as different cluster masses, densities, and density profiles. In order to perform a systematic and detailed study of how tidal shocks affect star clusters, the properties of both the model clusters and the shocks themselves have been idealised as much as possible to limit the number of free parameters influencing cluster evolution.

The $N$-body code \texttt{NBODY6tt} is a modified version of \texttt{NBODY6} \citep{aarseth03,aarseth10}, which models the external tidal field experienced by a stellar cluster according to a given tidal tensor. The main suite of models consists of clusters with initially $50,000$ stars of equal mass $0.6~\msun$ that are evolved in isolation for $1~\myr$. After that time, clusters are subjected to an extensive tidal shock of duration $\Delta t$, which is implemented by setting the first component of the tidal tensor to be positive, $T_{\rm xx} > 0$. Because we aim to study the effects of tidal shocks on cluster evolution throughout its lifetime, individual stars do not undergo stellar evolution. Cluster expansion due to stellar evolution is only an important factor at early times ($\la100$~Myr), but would need to be accounted for when applying tidal shock theory to very young star clusters.

The simulations are run with very small force calculation time-steps (of $\Delta t=$~0.003 Myr) to ensure the energy injected into each cluster by the tidal shock is modelled accurately. We summarize the characteristics of the models considered in Table~\ref{table:models}, with the names of the models describing the details of each cluster model and the tidal shock it undergoes. Specifically, model names reflect the shock strength (S), shock duration (L), cluster mass (M) and cluster density (D) relative to one another.

\begin{table*}
\centering
\begin{tabular}{lcccccc}%{llllll}
Name  & Shock Strength & Shock Duration & Initial mass & Initial half-mass radius & Initial density & Density Profile \\\hline
& $T_{\rm xx}\,[\myr^{-2}]$ & $\Delta t\,[\myr]$ & $M_{\rm i}\,[\msun]$ & $r_{\rm h,i}\,[\pc]$ & $\rho_{\rm h,i}\,[\msun \pc^{-3}]$ & \\\hline 
S1L1M1D27 & $5.75 \times 10^{-2}$ & 0.5 & 30,000 & 7.8 & 7.55 & Plummer\\
S1L2M1D27$^\star$ & $5.75 \times 10^{-2}$ & 1.0 & 30,000 & 7.8 & 7.55 & Plummer\\
S5L1M1D27 & $0.29$ & 0.5 & 30,000 & 7.8 & 7.55 & Plummer\\
S5L2M1D27$^\star$ & $0.29$ & 1.0 & 30,000 & 7.8 & 7.55 & Plummer\\
S15L1M1D27 & $0.89$ & 0.5 & 30,000 & 7.8 & 7.55 & Plummer\\
S15L2M1D27$^\star$ & $0.89$ & 1.0 & 30,000 & 7.8 & 7.55 & Plummer\\\hline
S5L1M02D27 & $0.29$ & 0.5 & 6,000 & 4.6 & 7.55 & Plummer\\
S5L2M02D27 & $0.29$ & 1.0 & 6,000 & 4.6 & 7.55 & Plummer\\
S5L1M2D27 & $0.29$ & 0.5 & 60,000 & 9.8 & 7.55 & Plummer\\
S5L2M2D27 & $0.29$ & 1.0 & 60,000 & 9.8 & 7.55 & Plummer\\
S5L1M1D729 & $0.29$ & 0.5 & 30,000 & 2.6 & 203.74 & Plummer\\
S5L2M1D729$^\star$ & $0.29$ & 1.0 & 30,000 & 2.6 & 203.74 & Plummer\\
S5L1M1D1 & $0.29$ & 0.5 & 30,000 & 23.0 & 0.29 & Plummer\\
S5L2M1D1$^\star$ & $0.29$ & 1.0 & 30,000 & 23.0 & 0.29 & Plummer\\
S5L1M1D27W5 & $0.29$ & 0.5 & 30,000 & 7.8 & 7.55 & King $W_0=5$\\
S5L2M1D27W5 & $0.29$ & 1.0 & 30,000 & 7.8 & 7.55 & King $W_0=5$\\
S5L1M1D27W7 & $0.29$ & 0.5 & 30,000 & 7.8 & 7.55 & King $W_0=7$\\
S5L2M1D27W7 & $0.29$ & 1.0 & 30,000 & 7.8 & 7.55 & King $W_0=7$ \\
\end{tabular}
\caption{\label{table:models}Summary of the models considered. The columns list the name of the model, the strength of the tidal shock, its duration, the initial cluster mass, the initial cluster half-mass radius, the volume density within the half-mass radius, and the initial density profile of the stellar cluster. The models in the first and second halves of the table list models with different tidal shock strengths and different cluster properties, respectively. The density profile of the cluster is only indicated in the name of the model if it differs from the default \citet{plummer11} profile. Stars next to the model name indicate that it is repeated for fifteen additional models that separate the shock into two sub-shocks of $0.5~\myr$ duration with different temporal offsets (see the text).}
\end{table*}

Firstly, to probe the dependence of the amount of mass lost on the tidal shock particularities, we perform simulations of clusters with the same initial density within the half-mass radius ($\rho_{\rm h,i} = 7.55~\msun~\pc^{-3}$, M1D27) that undergo a single tidal shock of different strengths, i.e.\ $T_{\rm xx} = \{5.75 \times 10^{-2}, 0.29, 0.86\}~\myr^{-2}$ (S1, S5, S15), and different durations, i.e.\ $\Delta t = \{0.5, 1\}~\myr$ (L1, L2). While the purpose of our study is to explore the effects that arbitrary tidal shocks have on star clusters with a range of properties, our choice of cluster and shock properties are motivated by physical processes. Specifically, the initial range of densities $\rho_{\rm h,i}$ is chosen to be comparable to high-density open clusters and low-density globular clusters \citep[e.g.][]{portegieszwart10,longmore14,krumholz19}. The ranges in shock strengths and durations are motivated by encounters between star clusters and GMCs. On the $\sim4$~pc size scale of the modelled clusters, the dynamical time for encounters with substructure in the interstellar medium is $0.2{-}5$~Myr \citep{efremov98,bolatto08,kruijssen13c}, implying that shock durations of 1~Myr accurately reflect encounters between star clusters and gas density peaks. Focusing on shock durations that are much less than the cluster's crossing time (4 Myr) has the added benefit of allowing us to consider impulsive shocks and thus ignore any correction factors accounting for the adiabatic expansion of the cluster.\footnote{We note that the relevant quantity of interest is not the shock duration itself, but the shock duration in units of the cluster crossing time, because that determines whether the shock is impulsive or drives an adiabatic response \citep[e.g.][]{gnedin97}.} Finally, the range of shock strengths that we consider ($T_{\rm xx}=0.06{-}0.86$~Myr$^{-2}$) corresponds to head-on (defined as occurring within $\sim10$~pc for typical GMC sizes) encounters with perturbers having central densities of $\rho_{\rm pert}\approx 2.3{-}34~\msun~\pc^{-3}\approx40{-}600~{\rm cm}^{-3}$, where the second equality assumes a perturber consisting of molecular gas. For more distant encounters, the density of the perturber scales as the impact parameter of the encounter cubed. Observed GMC densities range from $\rho_{\rm GMC}=10{-}10^4~{\rm cm}^{-3}$ \citep[e.g.][]{heyer09,longmore12}. The above examples demonstrate that the initial conditions adopted for the simulations presented here provide an accurate representation of tidal perturbations due to encounters with substructure in the interstellar medium, which are thought to dominate cluster disruption across cosmic time \citep[e.g.][]{gieles06,elmegreen10b,kruijssen11,miholics17}.

\begin{figure}
\centering
\includegraphics[width=0.5\textwidth]{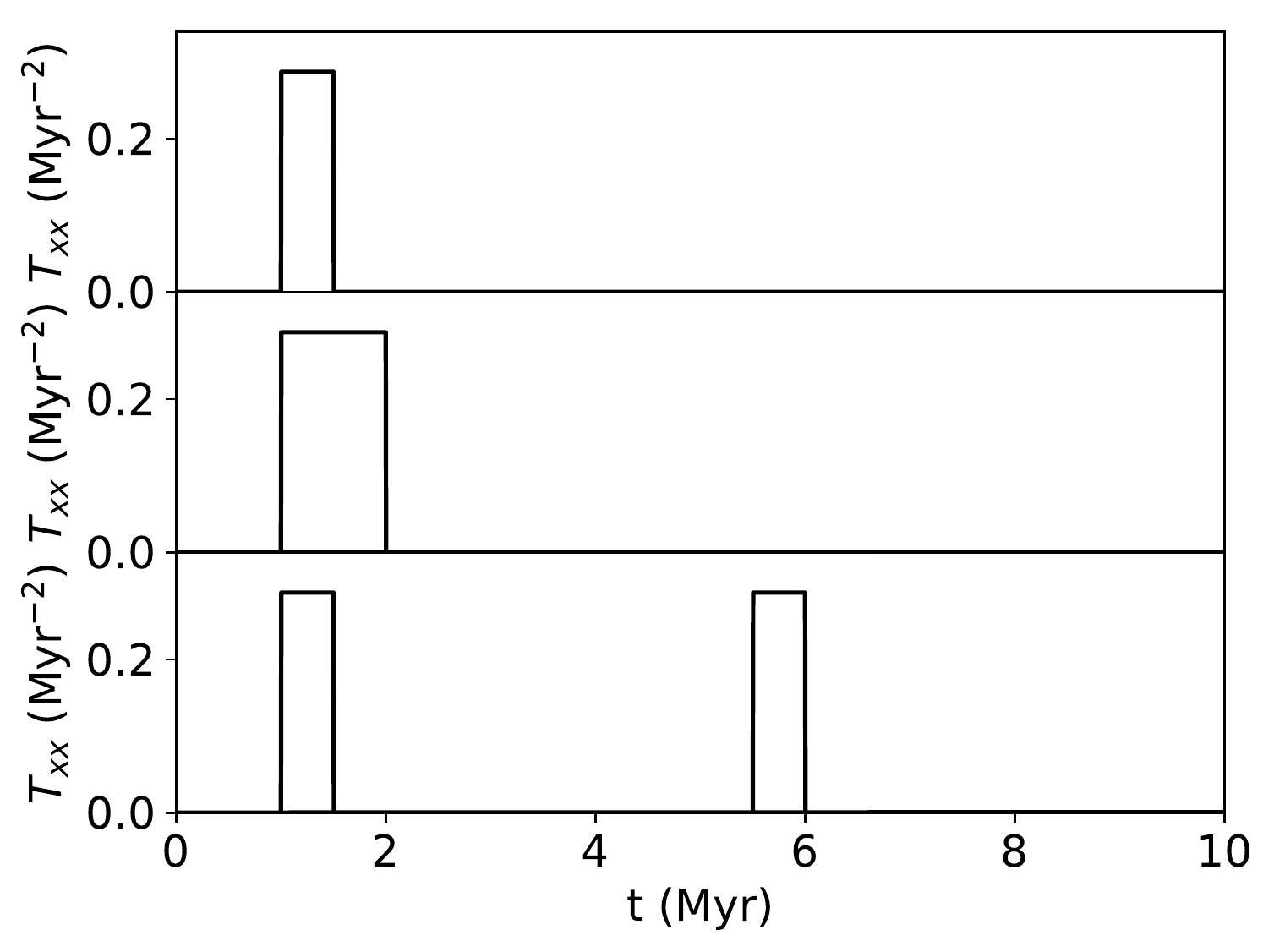}
\caption{\label{fig:txx} Time evolution of three tidal fields describing a tidal shock of strength $T_{\rm xx} = 0.29~\myr^{-2}$ with a duration of $\Delta t = 0.5~\myr$ (\textit{top}), $\Delta t = 1~\myr$ (\textit{middle}), and $\Delta t = 1~\myr$ with a gap of $t_{\rm gap} = 4~\myr$ between the first and the second half of the shock (\textit{bottom}).}
\end{figure}

We illustrate the tidal histories that we consider in Figure~\ref{fig:txx}, where we show the time evolution of the tidal tensor for a shock of strength $T_{\rm xx} = 0.29~\myr^{-2}$; the top and middle panels show the effect of changing the shock duration ($\Delta t = 0.5~\myr$ and $\Delta t=1~\myr$). These models represent our basic suite of simulations and are described in the first half of Table~\ref{table:models}.

Second, we explore how a cluster's response to a tidal shock depends on its own properties. To do so, we vary a single cluster property at a time: either their initial masses $M_{\rm i} = \{6, 60\}\times10^3~\msun$ (M02, M2), their initial densities within the half-mass radius $\rho_{\rm h,i} = \{0.29, 203.74\}~\msun~\pc^{-3}$ (D1, D729), or their initial density profiles. The lower and higher initial densities are comparable to the densities of typical open clusters and globular clusters respectively (see Figure 9 in \citealt{krumholz19}). All of these models are then evolved to undergo a tidal shock of strength $T_{\rm xx} = 0.29~\myr^{-2}$ and duration $\Delta t = \{0.5, 1\}~\myr$. For those models with different initial masses, their radii are adjusted to have the same initial density within the half-mass radius as our base models. For the initial density profile of the clusters, we consider King profiles \citep{king66} of parameters $W_0=\{5,7\}$ to compare them against the Plummer profiles used in our basic suite of models. This suite of models is described in the second half of Table~\ref{table:models}.

Finally, to explore how strongly the amount of mass loss depends on the exact shape of the tidal shock, we repeat models S1L2M1D27, S5L2M1D27,  S15L2M1D27 from Table~\ref{table:models}, i.e.\ stellar clusters of initial densities within the half-mass radii $\rho_{\rm h,i} = 7.55~\msun~\pc^{-3}$ that undergo a $1~\myr$ shock of strength $T_{\rm xx} = \{5.75\times 10^{-2}, 0.29, 0.89\}~\myr^{-2}$, respectively, with different tidal histories. We do the same for models with different initial densities, i.e.\ S5L2M1D1 and S5L2M1D729. This time, the shock experienced in the above seven models is split into two equal sub-shocks separated by different `gap' times $t_{\rm gap}$. We explore fifteen different gap times, of $t_{\rm gap} = \{0.25, 0.5, 1, 1.5, 2, 3, 4, 6, 8, 12, 16, 20, 24, 28, \allowbreak 32\}~\myr$, which includes gaps smaller and larger than the cluster's crossing time ($t_{\rm cr}\simeq 4~\myr$). The first component of the tidal tensor describing this type of shock with a gap time of $t_{\rm gap} = 4~\myr$ is illustrated in the bottom panel of Figure~\ref{fig:txx}. This suite of models is not included in the description of Table~\ref{table:models}, but they are noted in the legend and caption of any figures in which they are used.

\section{Evolution of the mass and the size of the cluster} \label{sec:results}

The immediate effect of a tidal shock on the evolution of a stellar cluster is to change its mass and size. During a tidal shock, the stars in the cluster gain energy \citep[e.g.][]{spitzer87}. Stars with sufficiently high energies will become unbound from the cluster and set the amount of mass lost. Less energetic stars will migrate outwards while remaining bound, thus driving the expansion of the cluster.

Using the models described in the previous section, we now discuss the evolution of the mass and the size of the cluster during a tidal shock for different tidal shock strengths, for shocks with different gap times and for clusters with different initial properties. In this discussion, we only consider stars that are energetically bound to each other to be members of the cluster.

\subsection{Dependence on the tidal shock strength}

In order to understand how the shock properties influence the amount of mass lost by a stellar cluster during a tidal shock, we first consider initially identical stellar clusters that are subject to tidal shocks of varying strength and duration (described in Section~\ref{sec:nbody} and summarised in the first half of Table~\ref{table:models}). 

\begin{figure*}
\centering
\includegraphics[width=\textwidth]{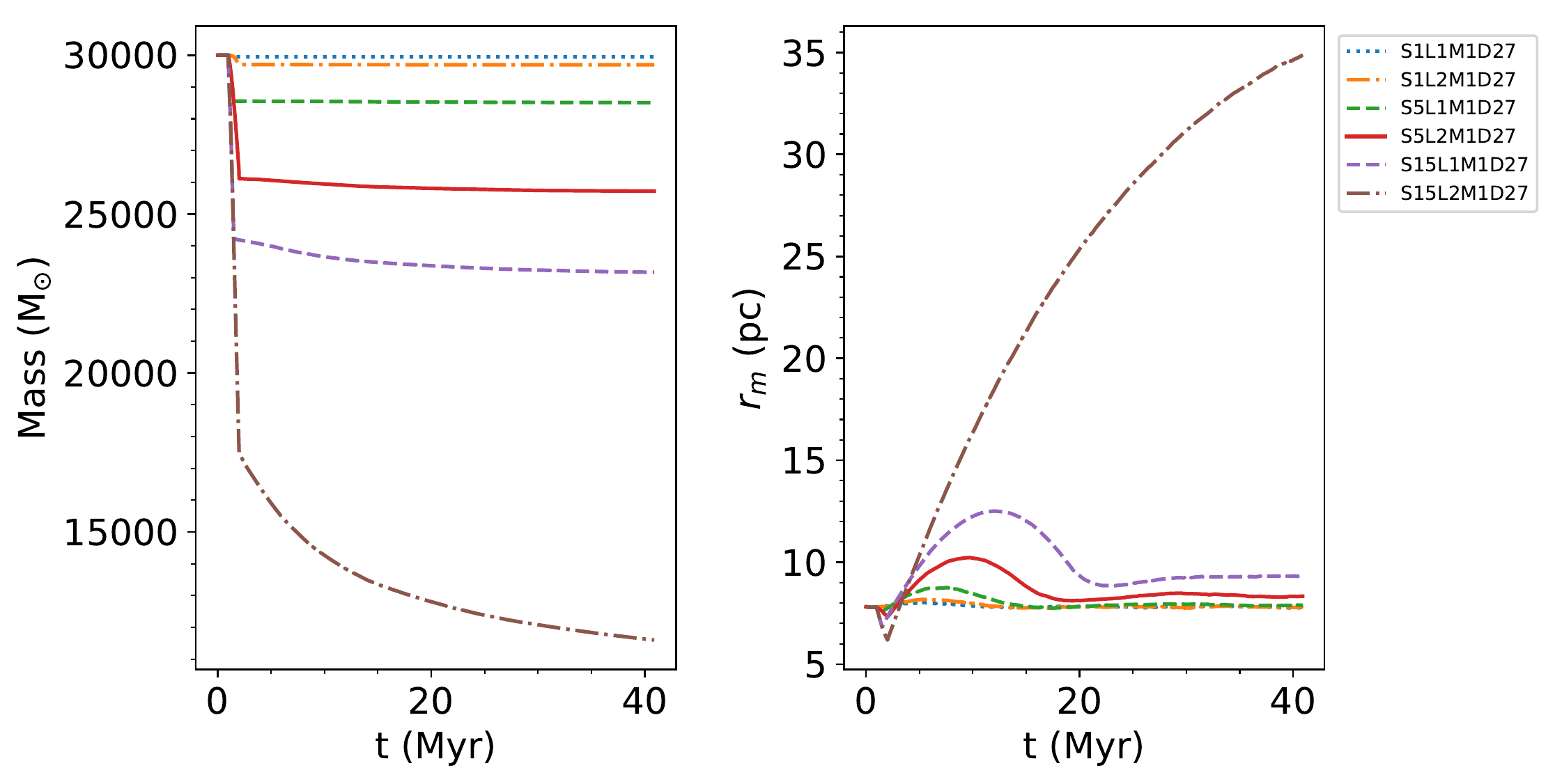}
\caption{\label{fig:mass-rm} Time evolution of the masses and half-mass radii of stellar clusters that undergo tidal shocks of different strength ($T_{\rm xx} = \{5.75\times10^{-2}, 0.29,0.89\}~\myr^{-2}$) and duration ($\Delta t = \{0.5,1\}~\myr$). The models are described in Section~\ref{sec:nbody} and summarised in the first half of Table~\ref{table:models}.}
\end{figure*}

The time evolution of the masses and the radii of the clusters are shown in the left and right-hand panels of Figure~\ref{fig:mass-rm}. We find that increasing the shock strength significantly affects the evolution of the cluster. Stronger tidal shocks, like the ones modelled in S15L1M1D27 and S15L2M1D27, inject more energy into the cluster, thus increasing both the amount of mass lost and the final radius of the cluster. The left-hand panel shows that the considered shock cause the clusters to lose $5$--$60$~per~cent of their initial mass, which increases with the integral of the shock and thus with its peak value and duration. The right-hand panel of Figure~\ref{fig:mass-rm} shows that, immediately after a tidal shock, the cluster rapidly shrinks. This occurs because the outer stars in the cluster experience the strongest energy gain and become unbound, leading to their rapid escape and reducing the half-mass radius of the stars remaining bound to the cluster. After the tidal shock, these remaining stars undergo a net expansion, as their gain in energy causes them to migrate outwards.

\subsection{Dependence on the shock interval}

We now explore how the mass loss depends on the distribution of the total amount of tidal heating over multiple shocks. To do so, we consider the idealised case of splitting the tidal shock into two sub-shocks of $0.5~\myr$ each, separated by a certain gap time $t_{\rm gap}$. Any configuration of these shocks corresponds to the same total amount of tidal heating, which is a function of the time integral of the tidal history, but the dynamical response of the cluster during the hiatus between the two sub-shocks might modify the amount of mass lost. To address this, we use models based on S5L2M1D27, but with different gap times as described in Section~\ref{sec:nbody}.

\begin{figure*}
\centering
\includegraphics[width=\textwidth]{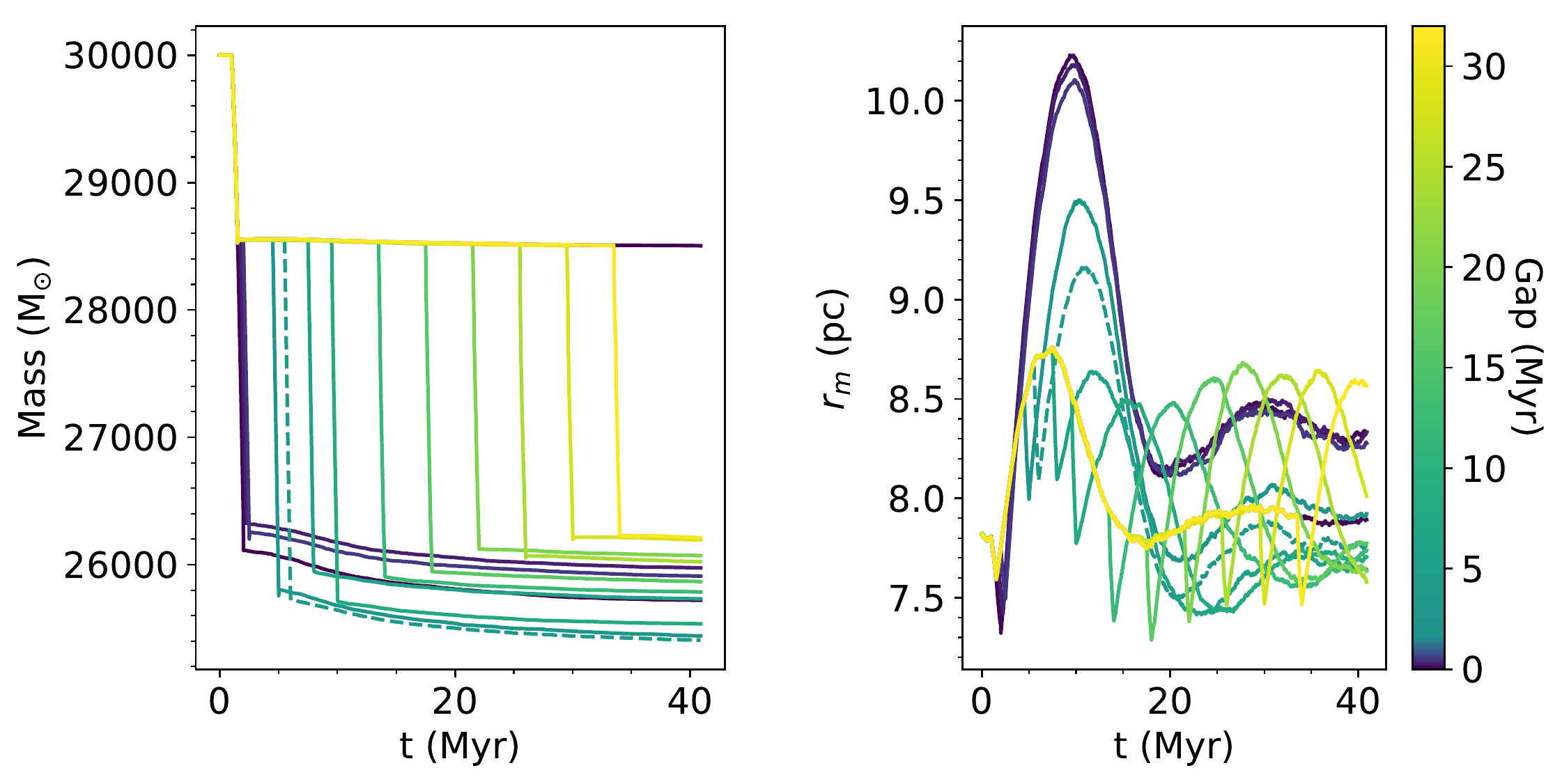}
\caption{\label{fig:gaptest} Time evolution of the masses and half-mass radii of stellar clusters that undergo two consecutive tidal sub-shocks of strength $T_{\rm xx} = 0.29~\myr^{-2}$ and a duration of $0.5~\myr$ each, which are separated by a certain time $t_{\rm gap}$ (indicated by the colours, with $t_{\rm gap}=4~\myr$ shown as a dashed line). These models are based on S5L2M1D27 and are described in Section~\ref{sec:nbody}.}
\end{figure*}

The resulting time evolution of the cluster masses and half-mass radii is shown in Figure~\ref{fig:gaptest}. Introducing a gap in the middle of the shock affects both the amount of mass lost and the radius evolution of the cluster. We find that those shocks with gap times smaller than the initial crossing time of the cluster, $t_{\rm gap}\la 4~\myr$, produce a negligible effect on the evolution of the mass and the size of the cluster, reaching comparable final masses, $M\simeq 2.6\times10^4~\msun$, and sizes, $r_{\rm h}\simeq 8.3~\pc$.

Conversely, models with tidal shocks separated by gap times longer than the cluster's initial crossing time ($t_{\rm gap} \ga 4~\myr$) exhibit a larger variety of final masses and sizes. In these models, the dynamical response of the cluster to the first sub-shock changes the mass and spatial structure of the cluster, such that it has a noticeably different mass, size and density profile by the time the second sub-shock begins. Hence, the mass loss and radius evolution caused by the second sub-shock depend on the new, evolved cluster properties, rather than on its initial mass and size. 

We also find that models with tidal shocks separated by gap times between $t_{\rm gap} = 8$--$16~\myr$ lose more mass during the second sub-shock than the models in which the sub-shocks are separated by $t_{\rm gap} < 8~\myr$. This difference in the amount of mass loss is caused by the considerable expansion of the cluster between the shocks for gap times $t_{\rm gap} = 8$--$16~\myr$, which implies a lower cluster density with a lower binding energy, making the stars more susceptible to becoming unbound during the second sub-shock. For models with gap times $t_{\rm gap} > 16~\myr$, the clusters have already contracted to near their original size by the time of the second sub-shock, such that their total mass loss is comparable to the models with gap times $t_{\rm gap} < 8~\myr$.

In summary, these results show that for tidal shocks separated over time-scales shorter than the initial crossing time of the cluster, fluctuations of (components of) the tidal tensor can be treated as a single tidal shock. However, fluctuations that are separated by a longer time interval should be considered independently when applying tidal shock theory to determine the induced mass loss and expansion.

\subsection{Dependence on the cluster properties}

Finally, we turn to the influence of the initial cluster properties on its response to a tidal shock. To do so, we consider models with different initial masses, half-mass radii, or density profiles, as described in Section~\ref{sec:nbody} and summarised in Table~\ref{table:models}. These models experience a tidal shock of strength $T_{\rm xx}=0.29~\myr^{-2}$ of two different durations ($\Delta t = \{0.5,1\}~\myr$).

\begin{figure*}
\centering
\includegraphics[width=\textwidth]{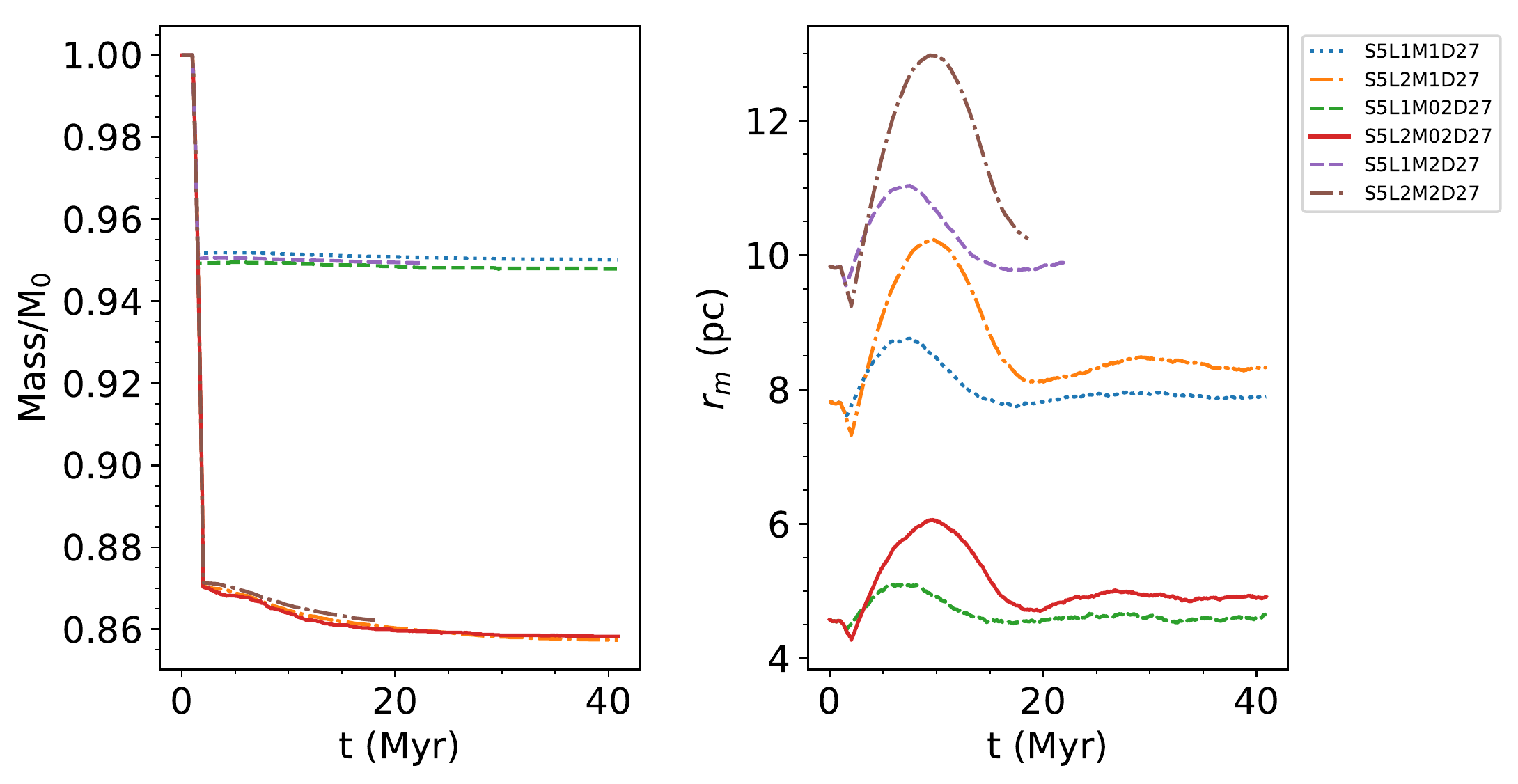}
\caption{\label{fig:mass-mi} Time evolution of the present-to-initial mass ratios of stellar clusters with different initial masses ($M_{\rm i} = \{6,30,60\}\times 10^3~\msun$), but the same volume densities, that undergo a shock of strength $T_{\rm xx}=0.29~\myr^{-2}$ of different durations ($\Delta t = \{0.5,1\}~\myr$).}
\end{figure*}

For cluster models with different initial masses, but identical densities as our basic models summarised in the first half of Table~\ref{table:models}, we show the time evolution of the present-to-initial mass ratio in Figure~\ref{fig:mass-mi}. The figure clearly shows that the response of the cluster to the tidal shock does not depend on its initial mass; the differences between some of the curves reflect different shock durations. Figure~\ref{fig:mass-mi} is consistent with the impulse approximation, which predicts that the change in mass is only dependent on cluster density and shock strength. Hence, we can focus solely on changing the initial densities of our model clusters to probe the effect that tidal shocks have on cluster evolution.

\begin{figure*}
\centering
\includegraphics[width=\textwidth]{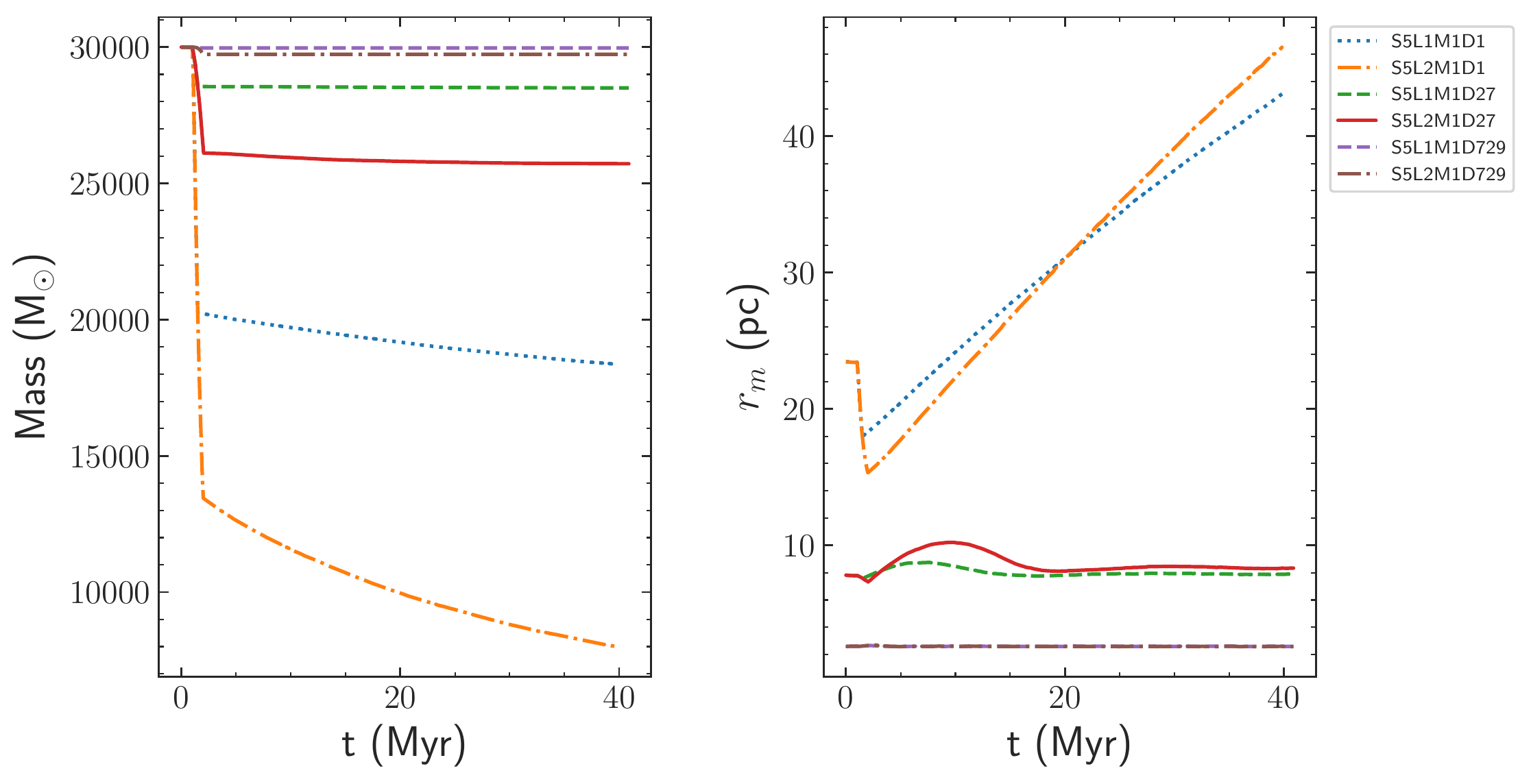}
\caption{\label{fig:mass-pi} Time evolution of the present-to-initial mass ratios of stellar clusters with different initial densities ($\rho_{\rm h,i}= \{0.29,7.55,203.74\}\times 10^3~\msun \pc^{-3}$) that undergo a shock of strength $T_{\rm xx}=0.29~\myr^{-2}$ of different durations ($\Delta t = \{0.5,1\}~\myr$).}
\end{figure*}

Figure~\ref{fig:mass-pi} shows clusters with the same properties as our basic set of models (see the first half of Table~\ref{table:models}), but with initial half-mass radii of $r_{\rm h,i} = \{2.6, 23\}~\pc$, and so, different initial densities within the half-mass radius, $\rho_{\rm h,i}= \{0.29,203.74\}\times 10^3~\msun \pc^{-3}$. As predicted, the mass loss strongly depends on the cluster density. This figure reveals that low-density models lose more mass than high-density models, by up to a factor of $\sim 4$ for densities differing by an order of magnitude. This dependence is consistent with stars having lower binding energies in the low-density models than in the high-density models, and thus, being more susceptible to becoming unbound and escaping from the cluster due to the energy gain from the tidal shock. The higher susceptibility of low-density clusters to tidal shocks is also reflected in the evolution of their half-mass radii. While high-density clusters undergo very little expansion after a shock, the low density models S5L1D27 and S5L2D27 expand by over a factor of~2 within $\sim40~\myr$, corresponding to 10 initial crossing times or 1-2 final crossing times.

Lastly, we consider the influence of the density profile of the cluster on its response to a tidal shock. We use the cluster models with the same properties as our basic set, but instead follow a King profile with parameters $W_0=\{5,7\}$ rather than the Plummer profile considered in the rest of this work. The shape of the profile affects the binding energies of the stars in the cluster. More concentrated profiles have more strongly bound stars that will experience less mass loss than extended profiles.

\begin{figure*}
\centering
\includegraphics[width=\textwidth]{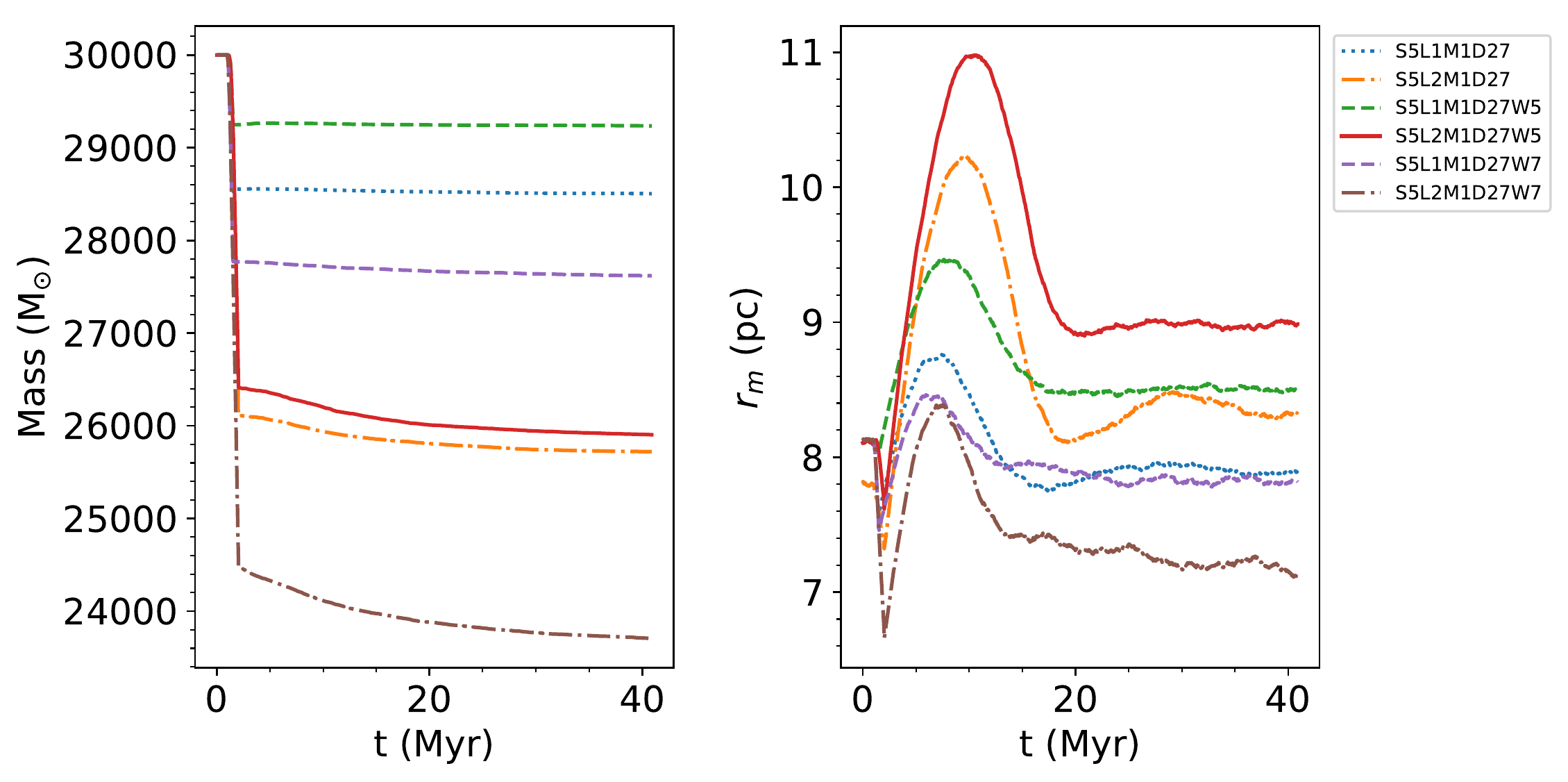}
\caption{\label{fig:mass-rm-king} Time evolution of the masses and half-mass radii of stellar clusters characterised by different density profiles (Plummer and King with $W_0=\{5,7\}$) that undergo a shock of strength $T_{\rm xx}=0.29~\myr^{-2}$ of different durations ($\Delta t = \{0.5,1\}~\myr$).}
\end{figure*}

The time evolution of the masses and half-mass radii of these models are shown in Figure~\ref{fig:mass-rm-king}. At a factor of 2 level, the differences in mass loss are subdominant relative to the density dependence shown in Figure~\ref{fig:mass-pi}. None the less, the models confirm that clusters with more extended density profiles lose more mass and expand more than centrally-concentrated clusters. These results are consistent with \citep{baumgardt03}, who observed a similar trend in model clusters undergoing shocks at perigalacticon.

In conclusion, the evolution of a stellar cluster during a tidal shock is most strongly affected by the strength and duration of the shock (i.e.\ the integral over the tidal tensor), as well by the cluster volume density. A second-order effect comes from the density profile, whereas the initial mass of the cluster has no impact on cluster mass loss. Across all presented models, fluctuations in the tidal tensor together constitute a single shock if they are separated by less than a crossing time, and should be considered as different shocks if they are separated by more than a crossing time.

\section{Discussion} \label{sec:discussion}

\subsection{Relative amount of mass lost in the $N$-body models}

A large body of literature has studied the effect of tidal shocks on stellar clusters from an analytical perspective \citep[e.g.][]{spitzer87,kundic95,gnedin99,gieles06,prieto08,kruijssen11,gieles16}. The common goal in these studies is to determine the tidal shock-driven mass loss due to various different sources. Most often, these consider the movement of a cluster in a galactic potential in which it experiences tidal shocks during the disc crossing and pericentre passage. Additional tidal histories considered include spiral arm crossings, GMC encounters, or any arbitrary history generated by the structure of the interstellar medium.

To calculate the mass loss due to tidal shocks, classical tidal shock theory uses the approach of determining how much energy is injected into a cluster over the duration of the shock, and relating the relative energy gain to the mass fraction that remains bound to the cluster after the shock \citep{spitzer58,spitzer87}. Given that the amount of mass lost is inversely proportional to the cluster disruption time-scale ($t_{\rm dis}$), if tidal shocks are the dominant mechanism driving cluster evolution, then the amount of mass lost is also inversely proportional to the shock-driven disruption time-scale ($t_{\rm sh}$), i.e.\
\begin{equation}
\dfrac{\Delta M}{\Delta t} = -\dfrac{M}{t_{\rm dis}} \rightarrow \dfrac{\Delta M}{M} = -\dfrac{\Delta t}{t_{\rm dis}} = -\dfrac{\Delta t}{t_{\rm sh}} .
\end{equation}
Using the tidal approximation, the shock-driven disruption time-scale in this expression can be related to the properties of the cluster and the tidal history \citep{prieto08}, i.e.\
\begin{equation}
t_{\rm sh} \propto \rho_{\rm h} I_{\rm tid}^{-1} = \rho_{\rm h} \left(\sum_{i,j} \left(\int T_{ij} dt\right)^{2}A_{{\rm w},ij}\right)^{-1},
\label{eqn:tsh}
\end{equation}
where $\rho_{\rm h}$ is the density of the cluster within the half-mass radius. The tidal heating parameter $I_{\rm tid}$ describes the increase in random motion (or heating) due to the tidal shock \citep{gnedin03}, and is determined as an integration over the duration of the shock of the tidal tensor components $T_{ij}$. Early work by \citet{weinberg94a,weinberg94b,weinberg94c} found that $T_{ij}$ needs to be corrected for energy loss due to the adiabatic expansion of the cluster, which later led to \citet{gnedin03} introducing the adiabatic correction terms $A_{{\rm w},ij}$. We can, however, omit this correction given that the impulsive shocks considered here are always much shorter than a crossing time. Assuming the change in cluster mass scales with the change in cluster energy, the shock-driven mass loss follows as
\begin{equation}
\dfrac{\Delta M}{M} =  -\dfrac{\Delta t}{t_{\rm sh}} \propto \rho_{\rm h}^{-1} \sum_{i,j}\left(\int T_{ij} dt\right)^{2} \Delta t .
\label{eqn:classic}
\end{equation}
In this equation, the mass loss depends on the integral of tidal tensor components over the course of the shock to the {\it second} power. This dependence makes it critical to properly identify individual shocks in order to correctly model their influence on the stellar cluster. For irregular tidal histories, this is highly non-trivial \citep[see e.g.][for discussions]{prieto08,kruijssen11,pfeffer18}, but our models with tidal shocks split into two sub-shocks with different gap times now enable a physically-motivated definition of what constitutes a single shock.

We now compare the amount of shock-driven mass loss in our models to the mass loss predicted by classical tidal field theory. In order to do so, we determine the amount of mass lost relative to the mass of the cluster before the shock, $\Delta M/M_{\rm pre-shock}$, for each of the models described in Section~\ref{sec:nbody}. For models that undergo a single shock, we determine the mass loss as the difference between the mass of the cluster at $t=40~\myr$, by when the shock-driven mass loss rate has become negligible, and the mass of the cluster when the shock begins at $t=1~\myr$, i.e.\ $\Delta M = M(40~\myr){-}M(1~\myr)$. However, for models where the shock is applied as two consecutive sub-shocks separated by a gap time $t_{\rm gap}$, we determine the mass lost during the two sub-shocks separately. The mass loss during the first sub-shock is always equal to the amount of mass lost by the model with a shock duration of $\Delta t = 0.5~\myr$, which is the same for all gap times. The mass loss during the second sub-shock is sensitive to the dynamical evolution of the cluster during the gap time. We use the mass of the cluster after the first sub-shock as its mass before the shock, $M_{\rm pre-shock} = M(1.5~\myr)$.

\begin{figure*}
\centering
\includegraphics[width=0.49\textwidth]{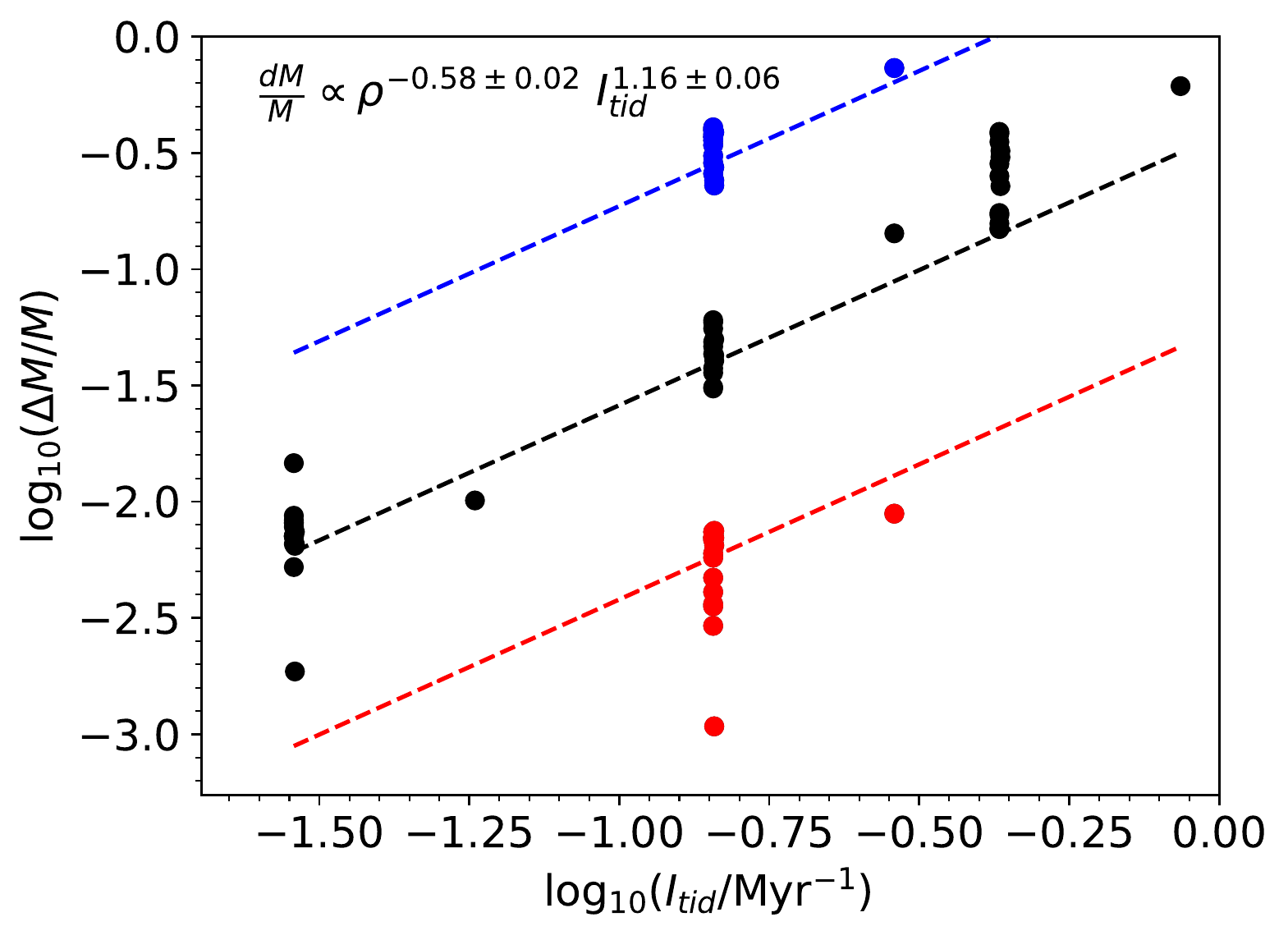}
\includegraphics[width=0.49\textwidth]{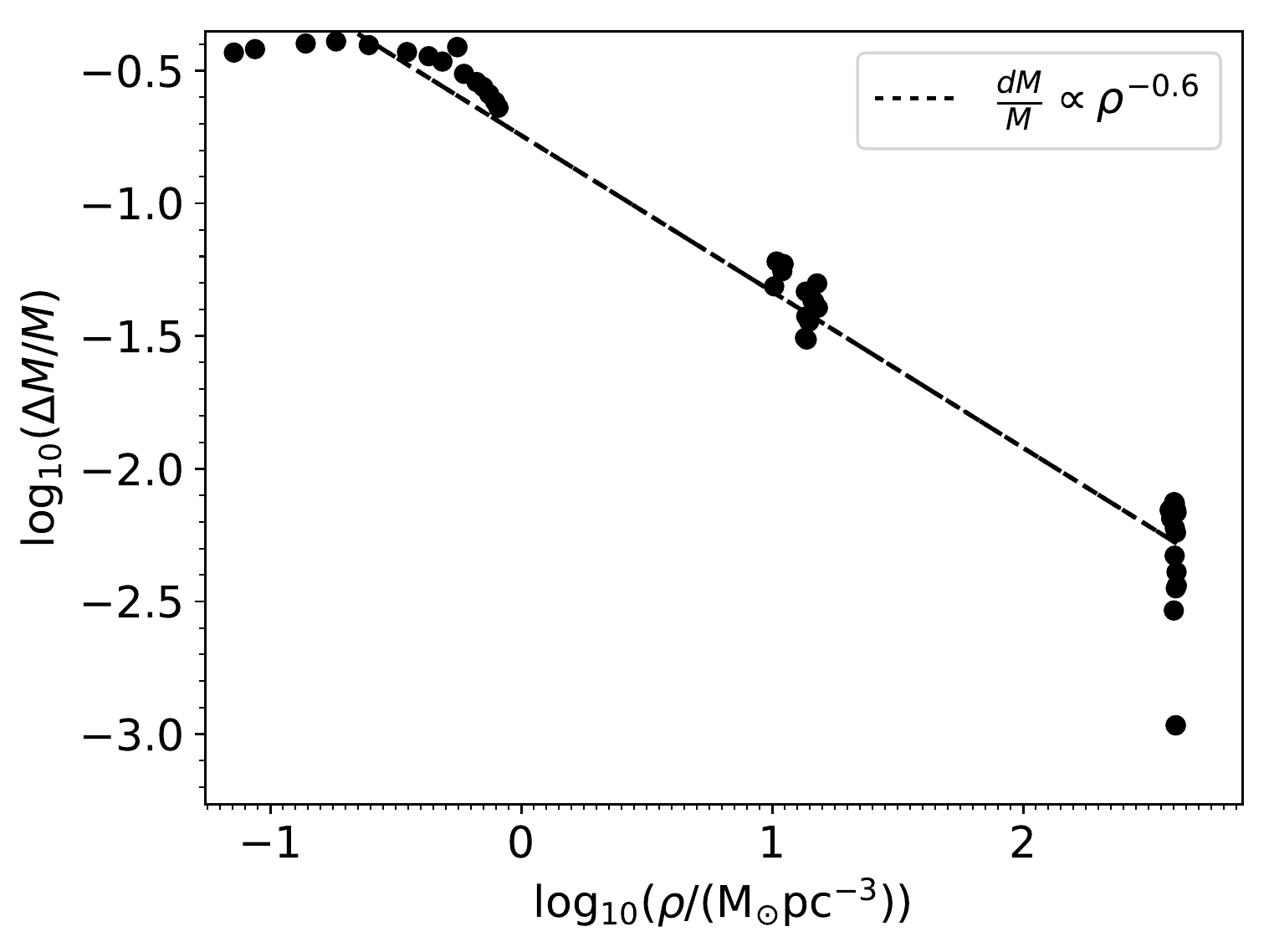}
\caption{\label{fig:ttplot}(\textit{Left}) Amount of mass loss in units of the pre-shock cluster mass as a function of the tidal heating parameter for the base models as well for the models with different initial densities and gap times, as described in Section~\ref{sec:nbody}. Models with the same initial half-mass density are shown in red, black and blue in order of decreasing density (D729, D27 and D1, respectively). The dashed lines visualise the best-fitting power law relation between the relative amount of mass loss, density, and tidal heating parameter (see the legend and the text). (\textit{Right}) Amount of mass loss in units of the pre-shock cluster mass as a function of the pre-shock half-mass density for the subset of models with different initial densities that are injected the same integrated tidal heating. The black dashed line indicates the best-fitting power law relation between the relative amount of mass loss and the density (see the legend and the text).}
\end{figure*}

For all of the models in the top half of Table~\ref{table:models} and the low and high density models that experience shocks of strength $T_{\rm xx}=0.29~\myr^{-2}$ and durations of $\Delta t = \{0.5,1\}~\myr$, the left panel of Figure~\ref{fig:ttplot} shows the amount of mass loss in units of the initial mass of the cluster before the shock, as a function of the tidal heating parameter. In all cases, models that experience two consecutive shocks (of duration 0.5~Myr) as opposed to a single shock (of duration 1~Myr) are also shown. For clusters that experience the same $I_{\rm tid}$, but have different initial densities, the amount of mass loss per unit pre-shock cluster mass is shown as a function of the initial half-mass density in the right panel of Figure~\ref{fig:ttplot}.

We perform a three-parameter power law fit to determine the relation between the relative amount of mass loss, the cluster density, and the tidal heating parameter as
\begin{equation}
\begin{split}
\MoveEqLeft
\dfrac{\Delta M}{M_{\rm pre-shock}} = A \left(\dfrac{\rho_{\rm h}}{10 \ \msun \pc^{-3}}\right)^{B} \left( \int \left(\dfrac{ T_{00}}{0.1 \ \myr^{-2}}\right) , dt\right)^{C}
%&\simeq 0.033 \left(\dfrac{\rho_{\rm h}}{10 \ \msun \pc^{-3}}\right)^{-0.59} \left(\int \left(\dfrac{ T_{00}}{0.1 \ \myr^{-2}}\right) dt\right)^{1.16} 
\end{split}
\label{eqn:fit}
\end{equation}
where we have simplified the tidal heating parameter to describe our extensive shock along the direction of the $x$-axis, without including the adiabatic correction. The best-fitting parameters are $A=0.033 \pm 0.007$, $B=-0.59 \pm 0.02$, and $C=1.16 \pm 0.06$, indicating that the results from our controlled $N$-body simulations differ significantly from the description considered in classical tidal field theory, for which $B=-1$ and $C=2$ (in agreement with \citealt{aguilar85}).

For a given initial half-mass density, the relative amount of mass lost depends almost linearly on the tidal heating parameter, $\Delta M/M_{\rm pre-shock} \propto I_{\rm tid}^{1.2}$. In addition, there is also a clear dependence on the density of the cluster -- for a given amount of tidal heating, lower density clusters lose more mass than their higher density counterparts (models R9 and R1, respectively). We quantify the dependence using the models from the second half of Table~\ref{table:models}, as they all experience a tidal shock that produces the same tidal heating. For these models, we find that the mass loss scales with the half-mass density as $\Delta M/M_{\rm pre-shock} \propto \rho_{\rm h}^{-0.6}$.

The above dependences break down for some of the lowest and highest density models, indicating there might be additional, second-order dependences. Both at the low and high-density ends, the models undergoing two sub-shocks separated by a long ($t_{\rm gap}=8{-}16~\myr$) gap time can expand substantially between the two-subshocks, such that their density profiles differ greatly from their initial profiles. The fact that our description breaks in these cases would indicate that their density profiles need to be taken into account as second order dependences, as suggested by Figure~\ref{fig:mass-rm-king}.

\begin{figure}
\centering
\includegraphics[width=0.48\textwidth]{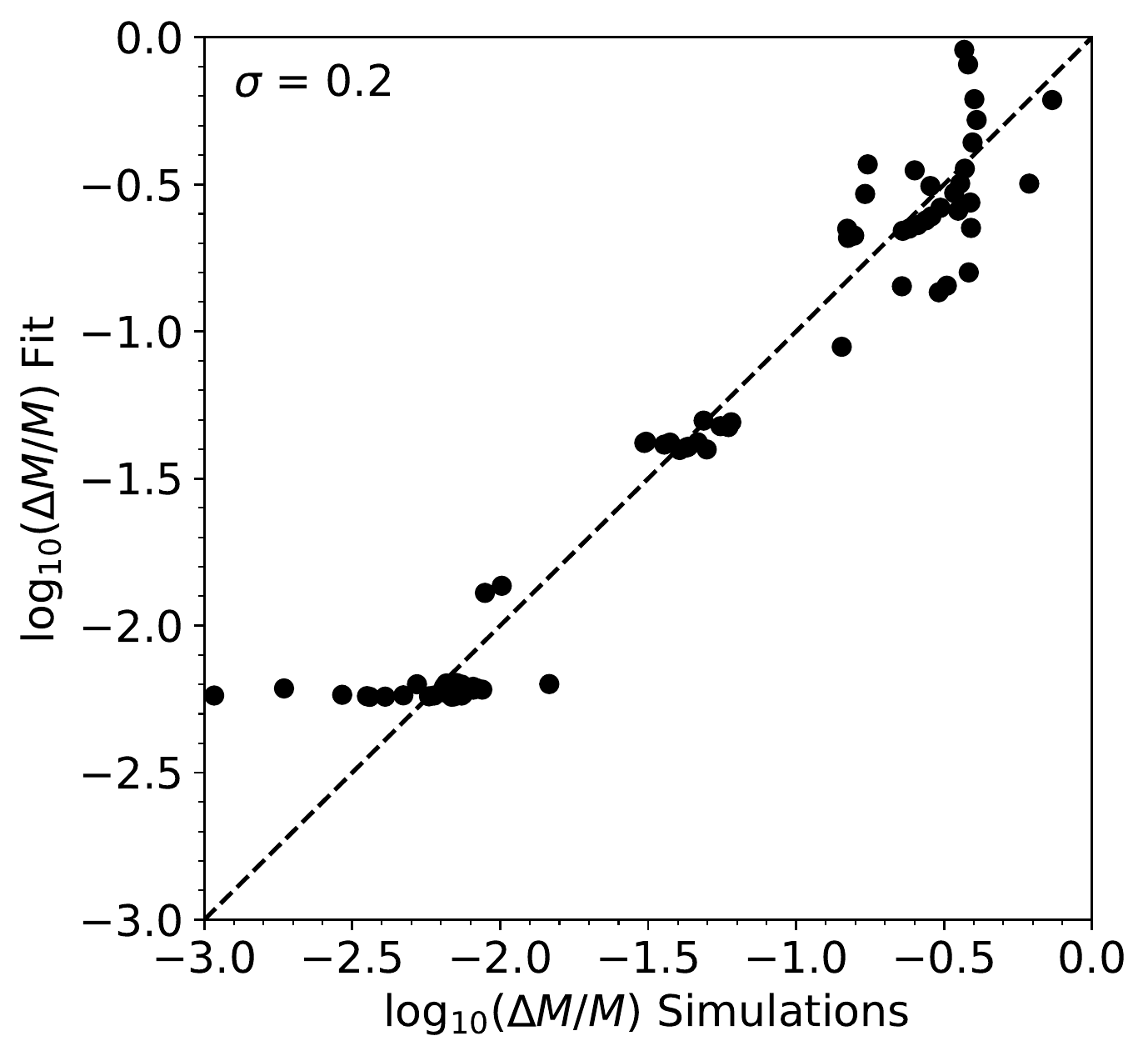}
\caption{\label{fig:match} Relative amount of mass lost determined from out fit in Eq.~(\ref{eqn:fit}) as a function of the relative amount of shock-driven mass loss experienced in our models. The standard deviation around the 1:1 (dashed) line is indicated in the top-left corner.}
\end{figure}

With the best-fitting description of the relative amount of mass loss, we can briefly explore how accurately it describes the shock-induced mass loss experienced by our models. Figure~\ref{fig:match} shows the relative amount of mass loss determined from Eq.~(\ref{eqn:fit}) as a function of the relative amount of mass loss found in our $N$-body models. The best fit exhibits a standard deviation of $\sigma=0.2$ relative to the results from the simulations. As discussed above, this scatter increases towards the larger and smaller relative amounts of mass loss, which is due to changes in the density profiles of those clusters, indicating that the amplitude of Eq.~(\ref{eqn:fit}) potentially hides a second-order dependence on the cluster density profile.

\subsection{A new theoretical model for tidal shocks}
The direct $N$-body simulations presented in this work show a clear discrepancy between the tidal shock-driven mass loss and the amount predicted by classical tidal shock theory. Comparing Eqs~(\ref{eqn:classic}) and (\ref{eqn:fit}), we find considerably shallower dependences on the tidal heating and the cluster half-mass density than previously suggested. This discrepancy indicates that the classical theoretical description cannot reproduce the true amount of mass loss across the full range of tidal heating parameters ($I_{\rm tid}=10^0{-}10^7~{\rm Gyr}^{-2}$, see e.g.\ \citealt{kruijssen11,pfeffer18}) and densities ($\rho=10^{-1}{-}10^5~\msun~\pc^{3}$, see \citealt{krumholz19}) expected across the full cluster population.

Importantly, classical tidal shock theory generally assumes that a star gets unbound from the cluster as soon as it attains a positive total energy. However, 
it was already proposed by \citet{chandrasekhar42} and followed up by \citet{king59} that stars that have gained total positive energy can remain bound if they encounter other stars on their way out of the cluster, such that they are re-captured by further two-body interactions that decrease their energy back to a total negative energy. \citet{king59} concludes that the retention of unbound stars that could potentially escape from an isolated cluster lengthens its lifetime. As previously mentioned, \citet{aguilar85} also pointed out that a star's ability to escape the cluster will also depend on the distribution of stellar velocities relative to the direction of the impulse provided by the tidal shock.

\citet{lee87} first considered the idea that during the evolution of a cluster in a static tidal field, stars that have gained sufficient energy from two-body interactions to leave the cluster still require some time to find the region around the Lagrange points where escape is actually achieved. The time-scale for escape therefore scales linearly with the crossing time. The inclusion of this escape time lengthens the lifetime of the cluster and is relevant both for isolated clusters and clusters in tidal fields \citep{baumgardt01}. The ability of individual stars to escape from the cluster depends on the distribution function of energies, which sets both the density profile and the velocity distribution of the stars; more extended clusters with higher-velocity stars will get disrupted sooner as they are more loosely bound.

We suggest to introduce a dependence of the disruption time due to tidal shocks on the escape time, analogously to the argument put forward by \citet{king59} and \citet{baumgardt01}. This accounts for the `back-scattering' of stars unbound by a tidal shock due to two-body encounters prior to their escape. The new description retains an influence on the shock strength and cluster density, as well as an additional dependence on the energy distribution function that characterises the density profile and the velocity of the stars in the cluster. Including an escape timescale should also alleviate the discrepancies found by \citet{aguilar85} between tidal shock theory and $N$-body simulations, as the distribution of stellar velocities is accounted for when estimating an escape timescale.

Following \citet{baumgardt01}, we consider two mechanisms that modify the energy of the stars in the cluster. Tidal shocks increase the energy of the stars, whereas two-body relaxation can provide or remove energy from the stars. The change of the distribution function of the energy [$N(\hat{E})$] with time is then given by
\begin{equation}
\dfrac{dN(\hat{E})}{d\hat{E}} = \dfrac{k_{1}}{t_{\rm rh}}\dfrac{d^2N(\hat{E})}{d\hat{E}^2} + \dfrac{k_{2}}{t_{\rm sh}}\dfrac{d^2N(\hat{E})}{d\hat{E}^2}
 - \hat{E}^2\dfrac{N(\hat{E})}{t_{\rm esc}},
\end{equation}
where $\hat{E} = (E-E_{\rm crit})/E_{\rm crit}$ and $E_{\rm crit}$ is the critical energy required to escape, $k_{1}$ and $k_{2}$ indicate how efficient each mechanism is at modifying the energy distribution of the cluster, and $t_{\rm rh}$, $t_{\rm sh}$ and $t_{\rm esc}$ are the half-mass relaxation, shock disruption and escape time-scales, respectively. Through this equation, we generalise the definition of the escape time-scale introduced by \citet{baumgardt01}, i.e.\ the time required to escape once a star attains positive energy, to also account for the energy input from tidal shocks.

Rearranging the previous equation and introducing the factor $\beta = k_1 t_{\rm sh}/k_2 t_{\rm rh}$, we find that the cluster disruption time-scale depends on the time-scales for shock disruption and escape as
\begin{equation}
t_{\rm dis} \propto {\left(\dfrac{t_{\rm sh}}{k_2}\dfrac{1}{1+\beta}\right)}^{3/4} {t_{\rm esc}}^{1/4}.
\label{eqn:th-cluster}
\end{equation}
The cluster disruption time takes on different forms depending on the dominant evolution mechanism. When tidal shock-driven disruption dominates, i.e.\ $k_1 t_{\rm sh}\ll k_2 t_{\rm rh}$ and $\beta\ll 1$, the disruption time depends solely on the shock time-scale as $t_{\rm dis} \propto {t_{\rm sh}}^{3/4} {t_{\rm esc}}^{1/4}$. However, when tidal shocks are sub-dominant, i.e.\ $k_1 t_{\rm sh}\gg k_2 t_{\rm rh}$ and $\beta\gg 1$, then we recover the solution described in \citet{baumgardt01}, where the disruption time depends on the half-mass relaxation time and the escape time-scale as $t_{\rm dis} \propto {t_{\rm rh}}^{3/4} {t_{\rm esc}}^{1/4}$.

\begin{figure}
\centering
\includegraphics[width=0.48\textwidth]{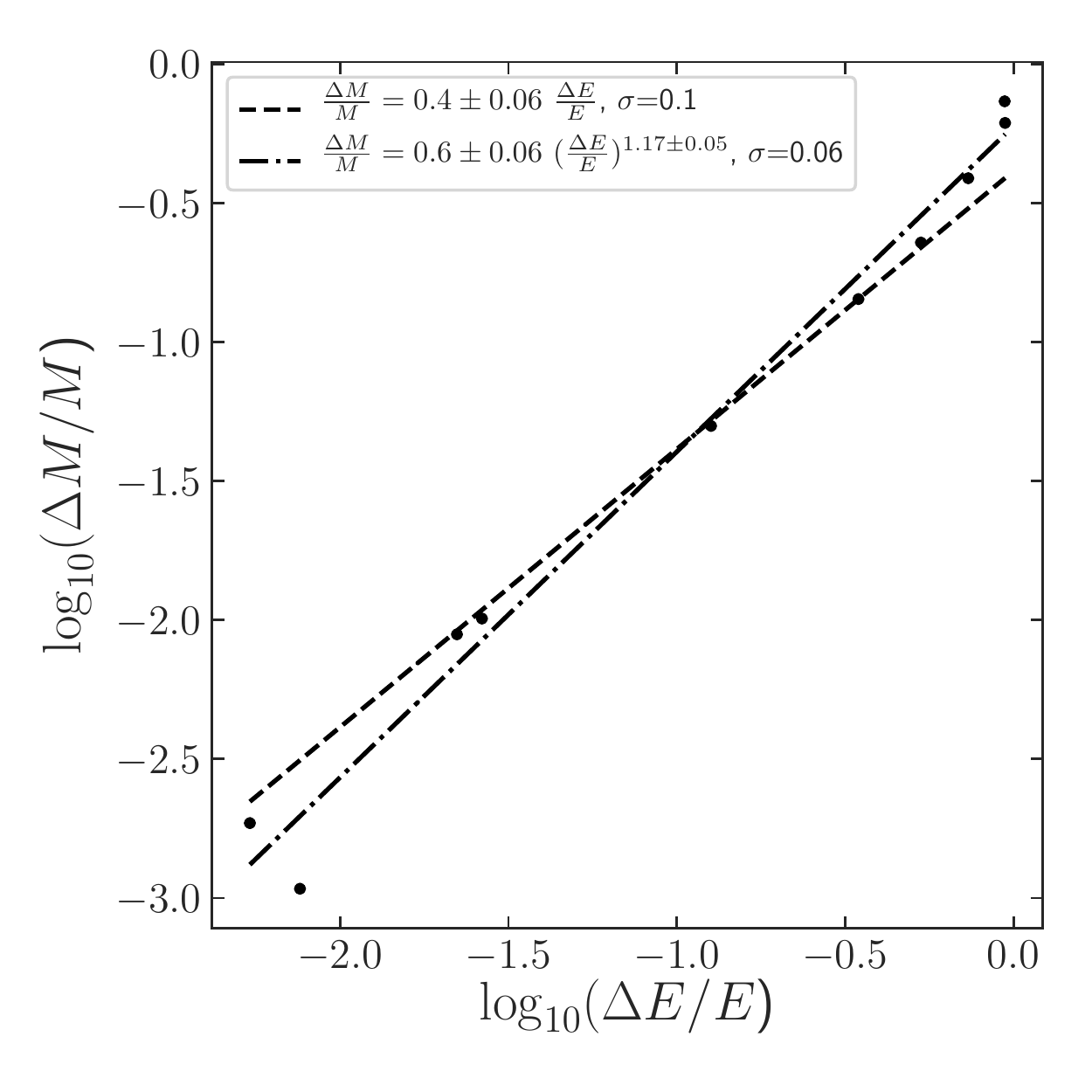}
\caption{\label{fig:deplot} Amount of mass loss in units of the pre-shock cluster mass as a function of the amount of energy loss in units of the pre-shock cluster binding energy for model clusters that undergo a single shock. The dashed line shows a fit to the data assuming $\Delta M/M$ and $\Delta E/E$ are linearly related (as classically assumed). The dash-dotted line shows a power law fit to the data, which assumes $\Delta M/M \propto (\Delta E/E)^{\alpha}$, with the best fit exponent being $\alpha=1.17 \pm 0.05$. In each case, the dispersion about the line of best fit is given.}
\end{figure}

As previously discussed, to relate a change in the energy of the cluster to a change in mass it must be assumed that $\Delta M/M = f \Delta E/E$. To test the validity of this assumption, we show in Figure~\ref{fig:deplot} how $\Delta M/M$ is related to $\Delta E/E$ in model clusters that undergo a single shock. Over the entire range of energies we find $f\sim0.4$ with a dispersion about the mean of $\sigma =0.1$. Alternatively, a power-law relation has a slightly lower dispersion than the linear case ($\sigma=0.06$), but has difficulty reproducing models with $\log_{10}(\Delta E/E) < -1.5$. Hence the assumption that $\Delta M/M$ and $\Delta E/E$ are linearly related is preferred over a power-law relation for $\log_{10}(\Delta E/E) < -0.2$. We therefore conclude that for the models considered here the best assumption is that $\Delta M/M = f \Delta E/E$. Nonetheless, we note that for strong tidal shocks or low density clusters, i.e.~in the regime of large changes in cluster energy, the assumption starts to break down and significant scatter will be introduced when attempting to predict changes in cluster mass. This finding is again in agreement with \citet{aguilar85}, who suggest that a single scaling relation between $\Delta M/M$ and $\Delta E/E$ may not exist over the entire range of $0 < \Delta E/E < 1$.

Using Eq.~(\ref{eqn:th-cluster}) to describe the disruption time, and assuming $\Delta M/M$ and $\Delta E/E$ are linearly related, we obtain that the amount of mass loss follows as
\begin{equation}
\dfrac{\Delta M}{M} = -\dfrac{\Delta t}{t_{\rm h}} \propto {\left(\dfrac{t_{\rm sh}}{k_2}\dfrac{1}{1+\beta}\right)}^{-3/4} {t_{\rm esc}}^{-1/4} \Delta t,
\label{eqn:dm-m-general}
\end{equation}
which implies a shallower dependence on the shock disruption time-scale than for the classical description in Eq.~(\ref{eqn:classic}). We define the escape time-scale to be proportional to the crossing time, $t_{\rm esc} \propto t_{\rm cr} = \sqrt{{r_{\rm h}}^3/G M}$, as it takes some multiple of a crossing time to change the orbital parameters of the stars in the cluster.

As first noted by \citet{aguilar85}, the amount of mass lost by a stellar population subjected to a tidal shock depends on the distribution of stellar velocities. The ability of an individual star to leave the cluster depends not only on their energy or their position within the cluster, but also on the orientation of its velocity relative to the principal direction of the tidal shock (given by the largest eigenvector of the tidal tensor, see \citealt{kruijssen11,pfeffer18}). Those stars with velocities parallel to the shock can escape more easily than those with velocities perpendicular or anti-parallel to the shock. In addition, tidal shocks modify the velocity field of the stars by introducing a degree of anisotropy along the principal direction of the shock, resulting in a higher probability of escape for those stars with velocities that are aligned with the shock.

\begin{figure}
\centering
\includegraphics[width=0.485\textwidth]{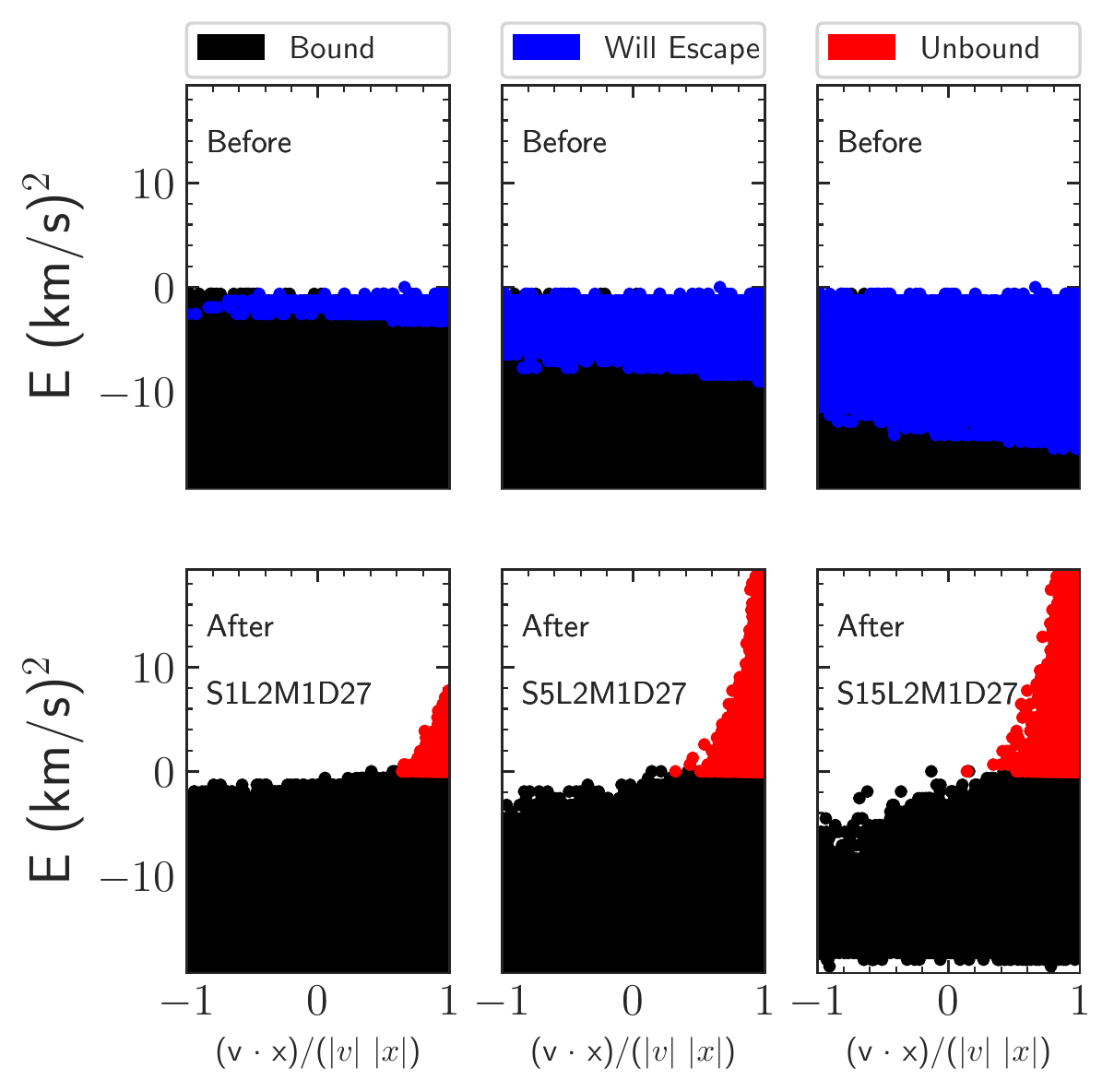}
\caption{\label{fig:evplot} Orbital energy as a function of the orientation between the main direction of the shock and the velocities of the stars before (top row) and after (bottom row) the shock for three models that describe shocks of different strength: S1L2M1D27 (\textit{left}), S5L2M1D27 (\textit{middle}) and S15L2M1D27 (\textit{right}). Bound stars are shown in black, stars that will become unbound due to the shock are marked in blue and unbound stars are shown in red.}
\end{figure}
Figure~\ref{fig:evplot} shows the orbital energy of stars before and after a shock, as a function of the orientation between their velocities and the principal direction of the shock, for three of our models that describe identical clusters that undergo shocks of different strength. Our models are initially characterised by a Plummer profile, so the clusters are spherically symmetric and the velocities of the stars are isotropic. This setup results in a uniform distribution of orientations between the velocities and the shock (top row in Figure~\ref{fig:evplot}). After the shock, the velocities become more parallel to its main direction, with the escaping stars having velocities that were predominantly oriented towards the direction of the shock before it occurred (bottom row in Figure~\ref{fig:evplot}). In order to account for the influence of the orientation between the velocities of the stars and the main direction of the shock in the ability of stars to escape, we consider the escape time-scale to be given by
\begin{equation}
t_{\rm esc} = \sqrt{\dfrac{{r_{\rm h}}^3}{G M}}(1-\mean{\cos(\theta)}) ,
\label{eqn:tescape}
\end{equation}
where $\mean{\cos(\theta)}$ is the mean cosine of the angle between the velocities of the stars and the principal direction of the shock. Our models are initially virialized and isotropic, so the mean cosine of the angle is $\mean{\cos(\theta)} = 0$ and the escaping time-scale is just the crossing time. However, if a cluster undergoes several shocks in the same direction without being able to virialise in between them, the velocity field will be more anisotropic and stars will be more likely to escape. In the more common case, where the distribution of repeated tidal shocks is isotropic, any anisotropy induced by a previous shock will have little influence over how a cluster responds to a second shock unless the time between shocks is short.

In order to quantify the degree of anisotropy introduced by a tidal shock in the velocities of the stars, we first need to determine the tidal force per unit mass experienced by the stars in the cluster. Following \citet{gnedin99} and assuming that stars do not move significantly during the tidal shock (i.e.~the impulse approximation), stars feel an acceleration ($\mathbf{a_{\rm sh}}$) due to the tidal encounter of
\begin{equation}
\mathbf{a_{\rm sh}} = -\left(\dfrac{\partial^2 \Phi}{\partial \mathbf{R}\partial \mathbf{R}} \right)\cdot \mathbf{x} ,
\end{equation}
where $\Phi$ is the potential describing the perturber causing the tidal shock, $\mathbf{R}$ is the position vector of the centre of the cluster relative to the perturber, and $\mathbf{x}$ is the position vector of a star relative to the centre of the cluster (note that vectors are indicated in boldface). We can rewrite this expression using the definition of the tidal tensor, ($T_{ij} = - \partial ^2 \Phi/ \partial R_j \partial R_i$) to obtain
\begin{equation}
\mathbf{a_{\rm sh}} = \sum_{j=0}^2 x_j \left(T_{0j}, T_{1j}, T_{2j}\right) ,
\label{eqn:ash}
\end{equation}
and the velocities of the stars after the shock are
\begin{equation}
\mathbf{v_{\rm sh}} = \mathbf{v} + \mathbf{\Delta v_{\rm sh}} = \mathbf{v} + \sum_{j=0}^2 x_j \left(\int T_{0j} {\rm d}t, \int T_{1j} {\rm d}t, \int T_{2j} {\rm d}t\right) ,
\label{eqn:vsh}
\end{equation}
where we integrate the tidal force per unit mass over the duration of the shock. 

The principal direction of the shock is given by the eigenvector corresponding to the largest eigenvalue of the tidal tensor, i.e.~a unit vector of the form $\mathbf{\hat{\lambda}} = (\lambda_{\rm x}, \lambda_{\rm y}, \lambda_{\rm z})/|\lambda|$. The anisotropy induced by the shock for a single star is then given by
\begin{equation}
\begin{split}
\MoveEqLeft
\cos(\theta) = \dfrac{\mathbf{\hat{\lambda}}\cdot \mathbf{v_{\rm sh}}}{|\mathbf{v_{\rm sh}}|} \\
& = \dfrac{\sum_{i=0}^2 \lambda_{\rm i} \left( v_{\rm i} + \sum_{j=0}^2 x_j \int T_{ij} {\rm d}t\right)}{\sqrt{|\mathbf{v}|^2 +2\sum_{i,j=0}^2 v_{\rm i} x_{\rm j} \int T_{ij} {\rm d}t + \sum_{i=0}^2\left(\sum_{j=0}^2 x_{\rm j} \int T_{ij} {\rm d}t \right)^2}}.
\end{split}
\end{equation}
The cluster is initially described by a density profile with a normalized distribution function ${\rm d}^2N/{\rm d}\mathbf{r} {\rm d}\mathbf{v}$. This allows us to integrate over phase space and determine the mean angle between the velocities of the stars and the main direction of the shock as
\begin{equation}
\begin{split}
\MoveEqLeft
\mean{\cos(\theta)} = \iint {\rm d}\mathbf{r} {\rm d}\mathbf{v} \dfrac{{\rm d}^2N}{{\rm d}\mathbf{r} {\rm d}\mathbf{v}}\\
&\times \dfrac{\sum_{i=0}^2 \lambda_{\rm i} \left( v_{\rm i} + \sum_{j=0}^2 x_j \int T_{ij} {\rm d}t\right)}{\sqrt{|\mathbf{v}|^2 +2\sum_{i,j=0}^2 v_{\rm i} x_{\rm j} \int T_{ij} {\rm d}t + \sum_{i=0}^2\left(\sum_{j=0}^2 x_{\rm j} \int T_{ij} {\rm d}t \right)^2}}.
\end{split}
\label{eqn:mean-costheta-general}
\end{equation}

In this work, we consider a single extensive tidal shock, such that a suitable choice of coordinates (with the shock acting along the $x$-axis) returns a tidal tensor that only contains a single positive component ($T_{00}$); all other components are zero. The principal direction vector is then given by $\hat{\lambda} =({\rm sgn}(x), 0, 0)$, where ${\rm sgn}(x) = x/|x|$ indicates the sign of the position of the star along the $x$-axis. As the shock only breaks the spherical symmetry for the $x$-axis, we can consider spherical coordinates rotated such that the spatial coordinates are described by $x=r\cos(\phi)$ and $R = r\sin(\phi) = \sqrt{y^2 + z^2}$ and the velocity coordinates are $v_{\rm x} = v \cos(\gamma)$ and $v_{\rm R} = v \sin(\gamma) = \sqrt{v_{\rm y}^2 + v_{\rm z}^2}$. Here, $\phi$ and $\gamma$ are the angles of the star's radius and velocity vectors relative to the $x$-axis (i.e.\ the principal direction of the shock), respectively, measured before the shock. With these definitions, we can rewrite Eq.~(\ref{eqn:mean-costheta-general}) as
\begin{equation}
\begin{split}
\MoveEqLeft
\mean{\cos(\theta)} = 4\pi^2\iint r^2\sin(\phi){\rm d}\phi {\rm d}r v^2\sin(\gamma) {\rm d}\gamma {\rm d}v \dfrac{{\rm d}^2N}{{\rm d}\mathbf{r} {\rm d}\mathbf{v}} \\
&\times \dfrac{{\rm sgn}[\cos(\phi)] \left[ v\cos(\gamma) + r \cos(\phi)\int T_{00} {\rm d}t \right]}{\sqrt{v^2 +2 v\cos(\gamma) r \cos(\phi) \int T_{00} {\rm d}t + \left[r \cos(\phi) \int T_{00} {\rm d}t\right]^2}} .
\end{split}
\label{eqn:mean-costheta-t00}
\end{equation}

\begin{figure*}
\centering
\includegraphics[width=0.99\hsize,keepaspectratio]{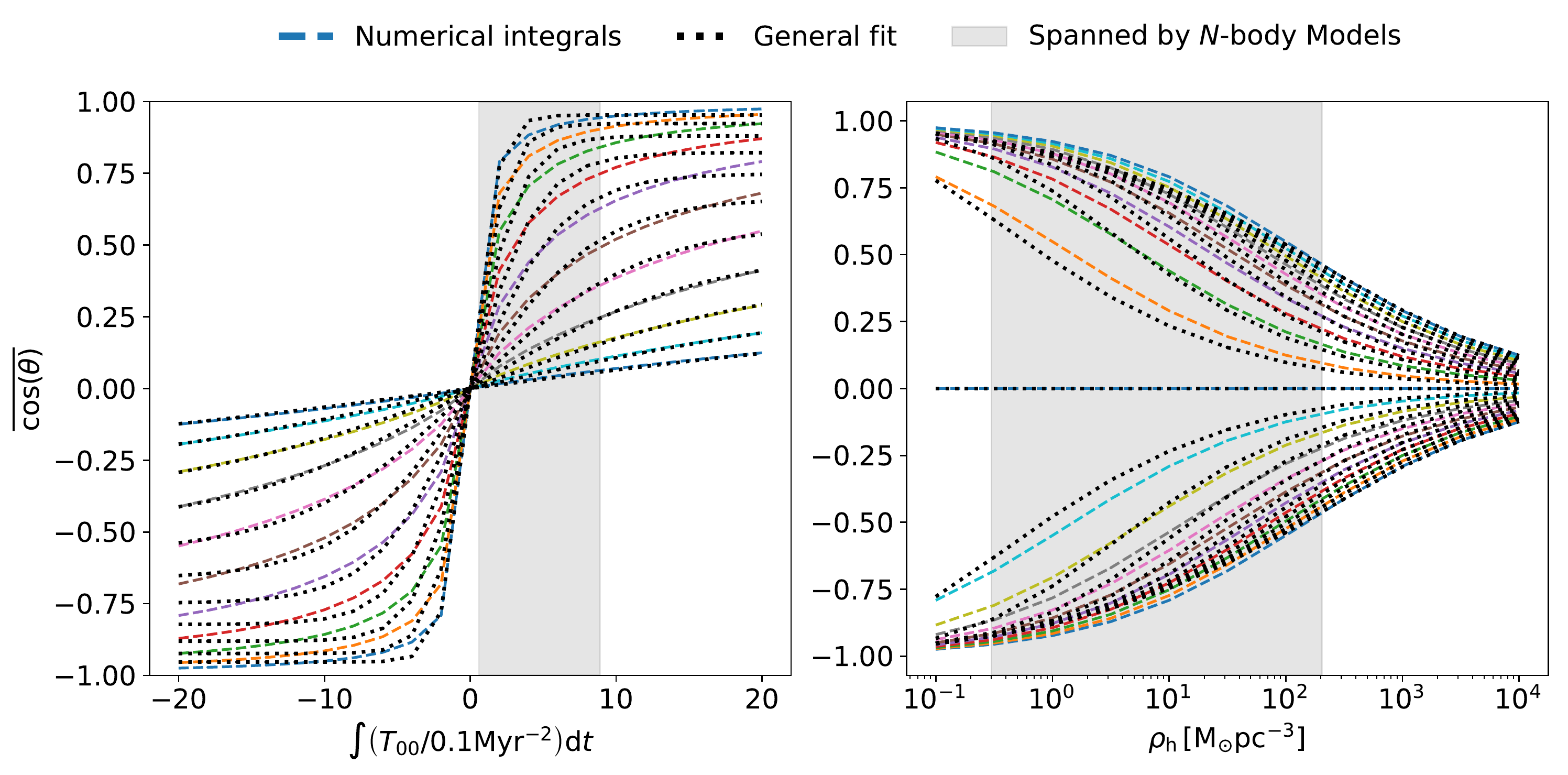}
\caption{\label{fig:prediction-cos-theta-fits} Mean cosine angle of the stellar velocity relative to the main direction of the shock as a function of shock strength (\textit{left}) and cluster half-mass density (\textit{right}). In the left and right panels, dashed lines represent the result of Eq.~(\ref{eqn:mean-costheta-plummer}) at various fixed values of the cluster half-mass density and shock strength, respectively. Black dotted lines represent the general fit described in Eq.~(\ref{eqn:cos-theta-genfit}). The range in shock strengths and cluster densities considered in this work is indicated by the grey shaded regions.}
\end{figure*}
\begin{figure}
\centering
\includegraphics[width=0.9\hsize,keepaspectratio]{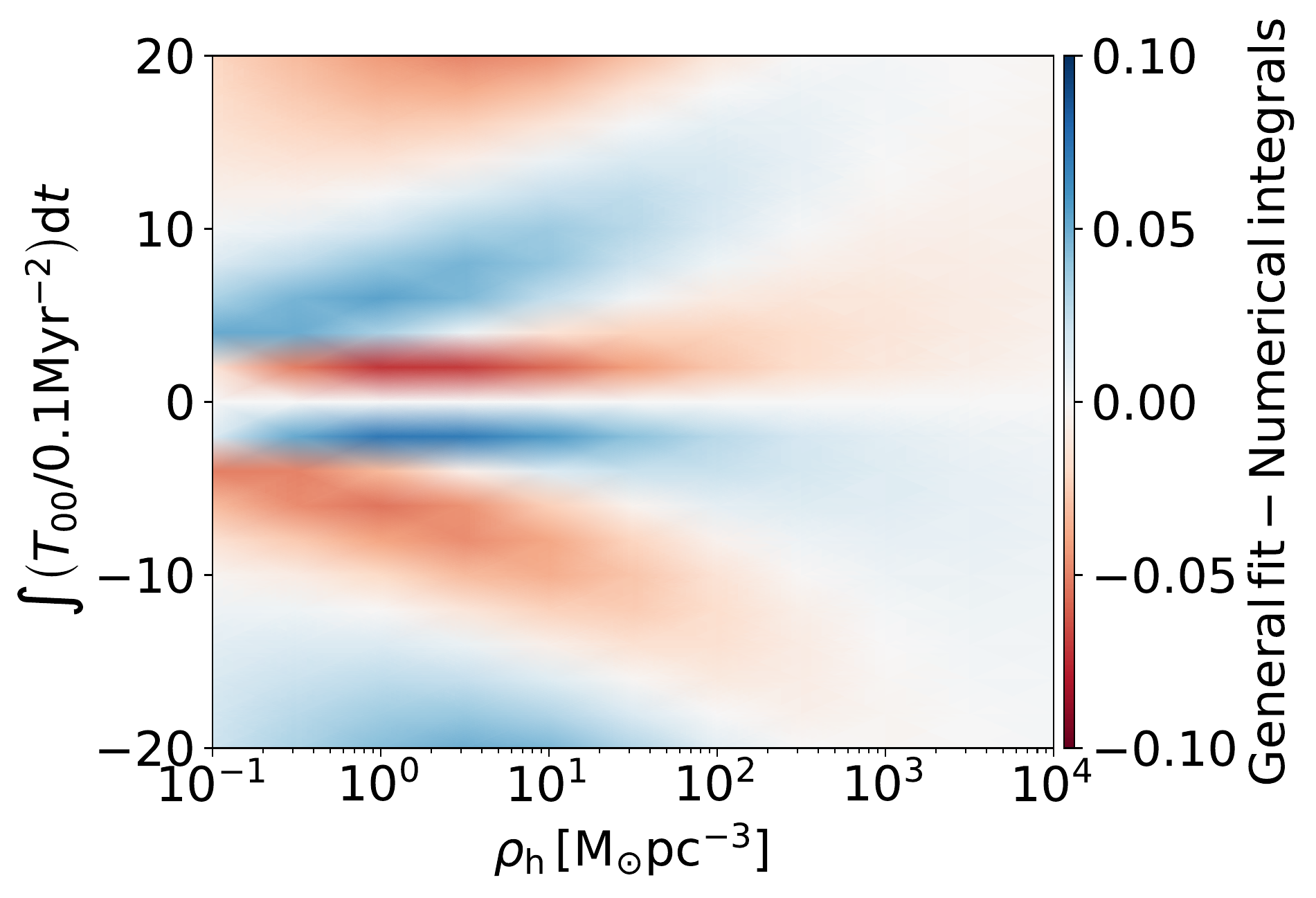}
\caption{\label{fig:prediction-cos-theta-diff} Absolute difference between the mean cosine angle of the stellar velocity relative to the main direction of the shock obtained as a result of Eq.~(\ref{eqn:mean-costheta-plummer}) and the general fit described in Eq.~(\ref{eqn:cos-theta-genfit}). The difference is $<0.1$ for all densities and shock strengths considered.}
\end{figure}

Assuming a Plummer profile, its normalised distribution function can be written in terms of the energy if the cluster is spherically symmetric and isotropic \citep{heggie03}, i.e.
\begin{equation}
\dfrac{{\rm d}N}{{\rm d}\mathbf{r} {\rm d}\mathbf{v}} = f(E) = \dfrac{3\times 2^{7/2}}{7\pi^3}\dfrac{a^2}{G^5 M^5}(-E)^{7/2} ,
\end{equation}
where $E = \Phi_{\rm P} (r) + v^2/2 = -GM/\sqrt{r^2 + a^2} + v^2/2$ is the total energy of a star in the cluster before the shock, $a\approx r_{\rm h}/1.3$ is the Plummer radius, $r_{\rm h}$ is the half-mass radius, $M$ is the cluster mass, and $G$ is the gravitational constant. Substituting this into Eq.~(\ref{eqn:mean-costheta-t00}), the mean cosine angle between the velocities and the shock after a tidal shock becomes
\begin{equation}
\begin{split}
\MoveEqLeft
\mean{\cos(\theta)} = \dfrac{12\times 2^{7/2}}{7\pi}\dfrac{a^2}{G^5 M^5}\\
&\times \int_{0}^{\pi} {\rm d}\phi \int_{0}^{\pi} {\rm d}\gamma \int_{0}^{\infty} {\rm d}r \int_{0}^{v_{\rm esc}(r)} {\rm d}v \sin(\phi)\sin(\gamma)r^2 v^2\\
&\times \dfrac{{\rm sgn}[\cos(\phi)] \left[ v\cos(\gamma) + r \cos(\phi)\int T_{00} {\rm d}t \right]}{\sqrt{v^2 +2 v\cos(\gamma) r \cos(\phi) \int T_{00} {\rm d}t + \left[r \cos(\phi) \int T_{00} {\rm d}t\right]^2}}\\
&\times \left(\dfrac{GM}{\sqrt{r^2 + a^2}} - \dfrac{1}{2}v^2 \right)^{7/2},
\end{split}
\label{eqn:mean-costheta-plummer}
\end{equation}
where $v_{\rm esc}(r) = \sqrt{-2\Phi_{\rm P}(r)} = \sqrt{2GM/(r^2+a^2)^{1/2}}$ is the escape velocity at radius $r$ of a Plummer sphere. By integrating up to $v_{\rm esc}(r)$, we are only including stars that remain bound to the cluster in our calculation of $\mean{\cos(\theta)}$ as the distribution function cancels for stars with $v=v_{\rm esc}(r)$ (i.e.~for those with null energy, $E=0$). If there is no tidal shock or if it is comparatively weak, the mean cosine angle after the shock will be similar to that before the shock, $\mean{\cos(\theta)} \simeq \mean{\cos(\gamma)} = 0$, where the equality to zero arises because we assume isotropy prior to the shock. To obtain a simple expression for the anisotropy introduced in the velocities by the shock, we can numerically integrate Eq.~(\ref{eqn:mean-costheta-plummer}) for different shocks strengths and cluster half-mass densities. We show the obtained mean cosine angles in Figure~\ref{fig:prediction-cos-theta-fits} and find that the mean cosine angle is well fitted by
\begin{equation}
\begin{split}
\MoveEqLeft
\mean{\cos(\theta)} \simeq A\left[1+\left(\dfrac{\rho_{\rm h}}{10~\msun\pc^{-3}}\right)^{B}\right]^{C}\\
&\times\tanh\left(D \left(\dfrac{\rho_{\rm h}}{10~\msun\pc^{-3}}\right)^E \int \left( \dfrac{ T_{00}}{0.1~\myr^{-2}}\right) {\rm d}t\right),
\end{split}
\label{eqn:cos-theta-genfit}
\end{equation}
% A = 1.00552821 +- 0.53692345
% B = 0.43666703 +- 1.23649459
% C = -0.42547109 +- 0.94918647
% D = 0.16149749 +- 0.12366816
% E = -0.27462901 +- 0.29695999
with $A=1.01\pm 0.54$, $B=0.44\pm1.24$, $C=-0.43\pm0.95$, $D=0.16\pm0.12$ and $E=-0.27\pm0.30$. We show the quality of the fit in Figure~\ref{fig:prediction-cos-theta-diff}. This type of dependence for the mean cosine angle indicates that the anisotropy in the velocity field of the cluster induced by a shock of a certain strength will depend on the cluster's half-mass density, with the anisotropy increasing towards lower densities for the same shock strength. At a given density, weak shocks will induce little anisotropy to the velocity field of the stars, but the anisotropy induced increases quickly with the shock strength, with a saturation limit (i.e.\ $\lim_{x\rightarrow0}\tanh{x}=1$) proportional to the cluster's half-mass density. Complete anisotropy, i.e. $\mean{\cos(\theta)}\approx1$, can only be induced for clusters with densities $\rho_{\rm h}\ll 1~\msun\pc^{-3}$ for the range of shock strengths considered in this work. 

We can combine the shock time-scale (Eq.~\ref{eqn:tsh}), the definition of the escape time (Eq.~\ref{eqn:tescape}), and the general fit for the mean cosine angle (Eq.~\ref{eqn:cos-theta-genfit}) to determine whether the amount of shock-induced mass loss in a cluster is significantly affected by the introduction of the escape time-scale. In the case of our extensive tidal shock along the $x$-axis, the shock time-scale and the escape time depend on the density of the cluster within the half-mass radius, the integral of the tidal shock strength and the mean cosine angle, $\mean{\cos(\theta)}$, as
\begin{equation}\label{eqn:new_model_eqn}
\begin{split}
\MoveEqLeft
\dfrac{\Delta M}{M} 
\propto t_{\rm sh}^{-3/4}  t_{\rm esc}^{-1/4} \\
&\propto \rho_{\rm h}^{-5/8} \left(\int T_{00} {\rm d}t\right)^{3/2} \left(1-\mean{\cos(\theta)}\right)^{-1/4}
\end{split}
\end{equation}

\begin{figure*}
\centering
\includegraphics[width=\textwidth]{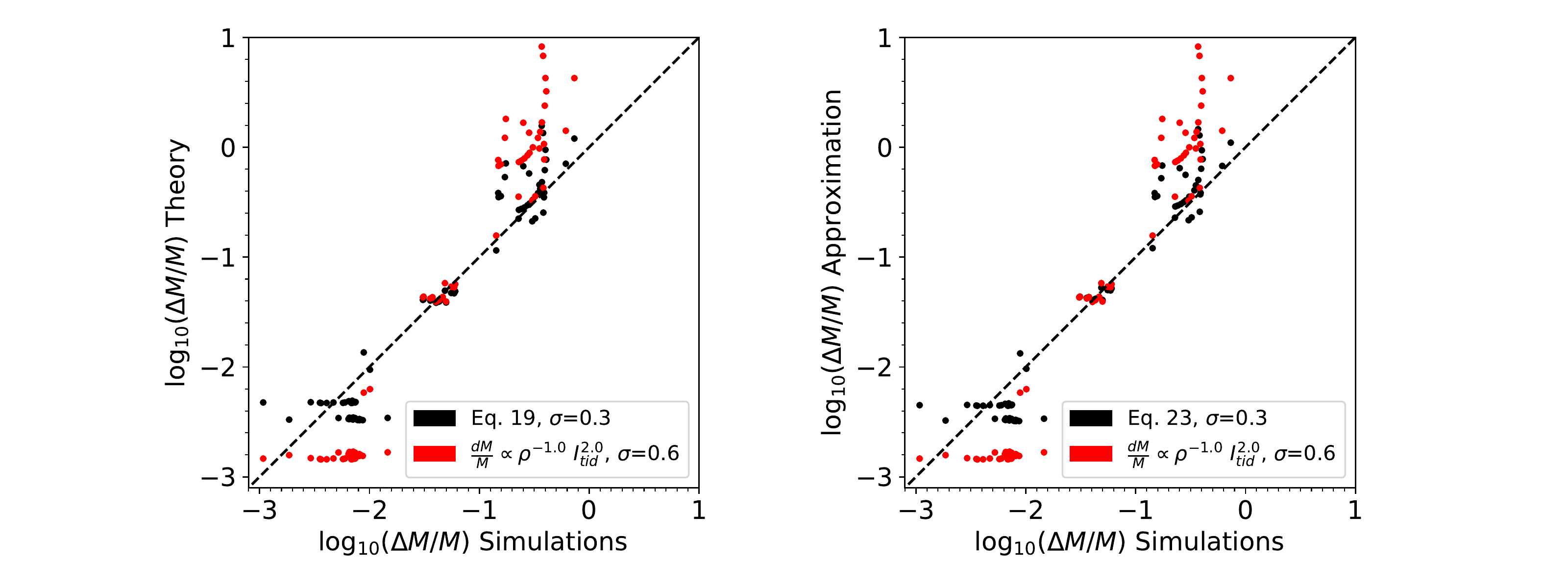}
%\hspace*{-0.05\textwidth}
%\includegraphics[width=0.49\textwidth]{compare_match_full.pdf}
%\includegraphics[width=0.49\textwidth]{compare_match_powerlaw.pdf}
\caption{\label{fig:compare_match} \textit{Left}: Relative amount of mass lost determined from our new theoretical model in Eq.~(\ref{eqn:new_model_eqn2}) (black) and classic tidal shock theory in Eq.~(\ref{eqn:classic}) (red) as a function of the relative amount of shock-driven mass loss experienced in our models. \textit{Right}: Same as the left panel, but with a power-law approximation in place of our new theoretical model (Eq.~\ref{eqn:new_model_powerlaw}). The standard deviation from the 1:1 dashed line is noted in the legend for each method. The new model including the escape time leads to a factor of 2 less scatter when predicting the relative amount of mass loss than classical tidal shock theory.}
\end{figure*}

Using the full suite of $N$-body simulations, we fit for the required proportionality constant in Equation~(\ref{eqn:new_model_eqn}), resulting in
\begin{equation}
\begin{split}
\MoveEqLeft
\dfrac{\Delta M}{M_{\rm pre-shock}} = A\,\left(\dfrac{\rho_{\rm h}}{10~\msun \pc^{-3}}\right)^{-5/8} \\
&\times\left(\int \left(\dfrac{ T_{00}}{0.1 \ \myr^{-2}}\right) {\rm d}t\right)^{3/2}\left(1-\mean{\cos(\theta)}\right)^{-1/4},
\end{split}
\label{eqn:new_model_eqn2}
\end{equation}
where the proportionality constant including its uncertainty is $A=2.06\pm0.04\times10^{-4}$. In the left panel of Figure~\ref{fig:compare_match}, we explore how well our new theoretical model estimates the relative amount of mass lost by each model cluster, in comparison to classical tidal shock theory. By accounting for the stellar escape time, this expression achieves a significant improvement in predicting the relative amount of tidal shock-driven mass loss compared to classic tidal shock theory. The new theoretical model (shown in black) predicts mass loss that is clustered around the 1:1 line, with a standard deviation around the line of $\sigma = 0.3$. This deviation is only slightly larger than the one obtained when carrying out a direct fit to the simulations (see Figure~\ref{fig:match} and Eq.~\ref{eqn:fit}). By contrast, classical tidal shock theory (shown in red) poorly predicts the relative amounts of mass lost by the simulations, resulting in a much larger standard deviation around the line of $\sigma = 0.6$. The largest discrepancy for classical tidal shock theory arises for both the weak (S1) and strong (S15) shock models and the low (D1) and high (D729) density models. These discrepancies are a direct result of overestimating the dependences on cluster density and tidal shock strength. We thus find that the introduction of the escape time is critical to accurately reproduce the response of shock-induced mass loss, as the discrepancy between predicted and observed mass loss is decreased by a factor of 2.

In order to further simplify the scalings of Eq.~(\ref{eqn:new_model_eqn2}), 
we restrict our analysis to the range of shock strengths and densities considered in this work (indicated in Figure~\ref{fig:prediction-cos-theta-fits} as grey shaded regions). Over this regime, we can approximate Eq.~(\ref{eqn:cos-theta-genfit}) in terms of a power-law, 
\begin{equation}
\begin{split}
\MoveEqLeft
1-\mean{\cos(\theta)} \propto \rho_{\rm h}^{B}\left(\int T_{00} {\rm d}t\right)^{C}
\end{split}
\label{eqn:cos-theta-powlaw}
\end{equation}
with $B\simeq 0.09$--$0.27$ and $C\simeq -0.52$ -- $0.00$ across the grey-shaded areas in Figure~\ref{fig:prediction-cos-theta-fits}. For this power-law approximation, the escape time follows as
\begin{equation}
\begin{aligned}
t_{\rm esc} &\propto \rho_{\rm h}^{B-1/2}\left(\int T_{00} {\rm d}t\right)^{C},
\end{aligned}
\end{equation}
with the power of the density now ranging between $B-1/2\simeq-0.41$-- $-0.23$. Substituting this expression along with Eq.~(\ref{eqn:tsh}) into Eq.~(\ref{eqn:th-cluster}) for the tidal shock-dominated case ($\beta\rightarrow0$), the amount of mass lost by a cluster during a tidal shock can be approximated as
\begin{equation}\label{eqn:new_model_eqn_powerlaw}
\begin{split}
\MoveEqLeft
\dfrac{\Delta M}{M} 
\propto t_{\rm sh}^{-3/4}  t_{\rm esc}^{-1/4} \\
&\propto \rho_{\rm h}^{-(5+2B)/8} \left(\int T_{00} {\rm d}t\right)^{(6-C)/4},
\end{split}
\end{equation}
with the slopes ranging between $-(5+2B)/8=-0.65$-- $-0.69$ and $(6-C)/4 = 1.5$--$1.63$. Adopting the mean slopes across these intervals, we can use the simulations to fit for the proportionality constant in Eq.~(\ref{eqn:new_model_eqn_powerlaw}), resulting in
\begin{equation}
\dfrac{\Delta M}{M_{\rm pre-shock}} = A_{\rm approx}\,\left(\dfrac{\rho_{\rm h}}{10 \ \msun \pc^{-3}}\right)^{-0.67} \left(\int \left(\dfrac{ T_{00}}{0.1 \ \myr^{-2}}\right) {\rm d}t\right)^{1.57} ,
\label{eqn:new_model_powerlaw}
\end{equation}
where the proportionality constant in the power-law approximation is $A_{\rm approx}=1.74\pm0.03\times10^{-4}$. As expected, the power-law approximation exhibits shallower dependences on the cluster density and the tidal shock strength than those predicted by classical field theory (Eq.~\ref{eqn:classic}). Comparing the exponential dependences in this result ($B_{\rm new}$,$C_{\rm new}$) to the fit to our $N$-body models from Eq.~(\ref{eqn:fit}) ($B_{\rm fit}$,$C_{\rm fit}$), the introduction of the escape time-scale reproduces the dependence on the density of the cluster (slopes of $B_{\rm new}=-0.67$ and $B_{\rm fit}=-0.59$, respectively), but predicts a somewhat steeper dependence for the tidal shock strength (slopes of $C_{\rm new}=1.57$ and $C_{\rm fit}=1.16$, respectively). Both represent a clear improvement relative to the classical description.

In the right-hand panel of Figure~\ref{fig:compare_match}, we explore how well the power-law approximation reproduces the relative amount of mass lost by each model cluster, again in comparison to classical tidal shock theory. We find that approximating the $1-\mean{\cos(\theta)}$ term as a power-law does very little to change our new theoretical model's ability to produce the fraction of mass lost by the model clusters. In fact, using the power-law approximation over the range in cluster densities and shock strengths covered by our models works extremely well, with the standard deviation around the one-to-one line in the right-hand panel of Figure~\ref{fig:compare_match} being $\sigma=0.3$ as in the left-hand panel. However, the power-law approximation is only valid for the range of parameters considered in this work. For a more general determination of the mean cosine angle, it is recommended to use the general fit described in Eq.~(\ref{eqn:cos-theta-genfit}) or solving the numerical integral described in Eq.~(\ref{eqn:mean-costheta-plummer}).

\section{Conclusions} \label{sec:conclusion}

We examine how stellar clusters evolve under the influence of tidal shocks in direct $N$-body simulations. We perform simulations of different model clusters subjected to evolving tidal tensors that represent different types of shocks. This approach allows us to study the influence of the properties of the shock and of the cluster on its mass and size evolution. 

We find that clusters undergoing a single tidal shock react differently depending on the strength of the shock. Over a factor of 15 in shock strength, we find a factor of $\sim60$ variation in shock-induced mass loss, with stronger shocks leading to more mass loss. When the tidal shock is applied as two sub-shocks separated by a certain gap time, the mass lost during the second sub-shock is sensitive to the dynamical evolution of the cluster during the gap time. Those clusters that encounter the second sub-shock faster than their crossing time ($t_{\rm gap}\la4~\myr$) exhibit a similar evolution of their mass and radius by the end of the simulation. By contrast, the clusters that encounter the second sub-shock on time-scales longer than their crossing time ($t_{\rm gap}\ga4~\myr$) exhibit a large variation of their final masses and sizes, indicating that their evolution in between both sub-shocks affects how they respond to the second sub-shock.

The response of the models to single and consecutive shocks indicates that, within one crossing time of the cluster, a tidal shock should be defined as a single, uninterrupted injection of energy with a magnitude equal to the integral of the tidal tensor. Likewise, the cluster density can be evaluated at the beginning of the shock. Over time-scales longer than the cluster's crossing time, energy injections should be treated as multiple tidal shock events. The appropriate cluster density must be evaluated at each of these individual events.

The initial mass of the cluster does not have a major effect on the response of the cluster to the tidal shock, but the cluster's evolution is controlled by its density within the half-mass radius. Across a factor of $\sim700$ in density, the shock-induced mass loss varies by up to a factor of $80$, with lower-density clusters losing less mass. This happens because the stars in lower-density clusters are more loosely bound and can escape more easily. We identify a second-order dependence on the cluster density profile, expressed in terms of the King parameter $W_0$, such that less concentrated cluster profiles exhibit elevated mass loss.

We determine the mass loss in units of the initial cluster mass for all of our models, and find that it depends on the tidal shock strength and the cluster density as $\Delta M/M\propto\rho^{-0.59} \left(\int T_{00} {\rm d}t \right)^{1.16}$. These dependences are considerably shallower than those predicted by classical tidal shock theory, which predicts $\Delta M/M\propto\rho^{-1} \left(\int T_{00} {\rm d}t \right)^{2}$ when using the tidal approximation. In order to explain this discrepancy, we propose that stars require a certain time to escape the cluster once their energy has become positive due to the shock, analogously to the model of \citet{baumgardt01} for mass loss driven by two-body relaxation. We define an escape time-scale proportional to the crossing time and a factor accounting for the orientation of the velocities of the stars in the cluster relative to the principal direction of the shock. The introduction of this time-scale lengthens the cluster's lifetime and correctly predicts the amount of shock-induced mass loss over a range of shock strengths and cluster densities via Eq.~(\ref{eqn:new_model_eqn2}). The relation is an accurate method for predicting mass loss due to tidal shocks, while still using the tidal approximation, instead of having to know the details of each tidal shock event (i.e. the density profile and orbit of the perturber). Approximating this new formalism for shock-induced mass loss as a power-law for the range of shock strengths and densities considered in this work, we obtain $\Delta M/M\propto\rho^{-0.67} \left(\int T_{00} {\rm d}t \right)^{1.57}$, which is also in considerably better agreement with our simulations than the classical prediction.

Incorporating a dependence on the stellar escape time into tidal shock theory significantly improves our ability to predict the shock-induced mass loss. This new model is able to correctly predict the relative amounts of mass loss ($\Delta M/M$) found in the simulations with a standard deviation of $\sigma = 0.3$. This amount of scatter is similar to that obtained when fitting the results of our simulations for the power law dependence on density and shock strength ($\sigma=0.2$). Hence, the new model provides an excellent description of the simulations. Without a dependence on the escape time, we instead obtain a dispersion of $\sigma = 0.6$ and are unable to retrieve the absolute amount of mass loss over the full range of model cluster densities and tidal shock strengths. Our new model also accounts for the identified second-order dependence on the cluster density profile. In order to determine the degree of anisotropy of the stellar velocity distribution induced by the tidal shock, we consider a Plummer profile (Eq.~\ref{eqn:mean-costheta-plummer}), but a similar derivation for a King model could be easily done by replacing the appropriate description in Eq.~(\ref{eqn:mean-costheta-t00}).

Developing an accurate model for the impact of tidal shocks on the evolution of stellar clusters is an important step, because studies of cluster evolution continue to move away from static, smooth potentials, and instead model the evolution of clusters in a cosmological context. Over the course of a cluster's lifetime (and especially at early times), tidal shocks represent the dominant source of dynamical mass loss. Therefore, being able to predict shock-driven mass loss rate is critical for describing the formation and evolution of stellar cluster populations. By accounting for shock-driven mass loss in a cosmological context \citep[as is now being pursued, see][]{pfeffer18}, the present-day properties of globular cluster populations may be connected to those at the time of their formation. In turn, this will enable the use of globular clusters as direct tracers of the formation and evolution of their host galaxy \citep[e.g.][]{kruijssen19c}.

\section*{Acknowledgements}

JW acknowledges financial support through a Natural Sciences and Engineering Research Council of Canada (NSERC) Postdoctoral Fellowship and additional financial support from NSERC (funding reference number RGPIN-2015-05235) and an Ontario Early Researcher Award (ER16-12-061). MRC is supported by a Fellowship from the International Max Planck Research School for Astronomy and Cosmic Physics at the University of Heidelberg (IMPRS-HD). MRC and JMDK gratefully acknowledge funding from the European Research Council (ERC) under the European Union's Horizon 2020 research and innovation programme via the ERC Starting Grant MUSTANG (grant agreement number 714907). JMDK gratefully acknowledges funding from the German Research Foundation (DFG) in the form of an Emmy Noether Research Group (grant number KR4801/1-1). This work was made possible in part by the facilities of the Shared Hierarchical Academic Research Computing Network (SHARCNET: \url{www.sharcnet.ca}) and Compute/Calcul Canada, in part by Lilly Endowment, Inc., through its support for the Indiana University Pervasive Technology Institute, and in part by the Indiana METACyt Initiative. The Indiana METACyt Initiative at IU is also supported in part by Lilly Endowment, Inc. We thank Gonzalo Alonso-\'Alvarez, Sarah Jeffreson, and Benjamin Keller for helpful discussions, as well as Mark Gieles, Oleg Gnedin, and Simon White for feedback on the original version of this paper.

%%%%%%%%%%%%%%%%%%%%%%%%%%%%%%%%%%%%%%%%%%%%%%%%%%

\bibliographystyle{mnras}
\bibliography{bibfile} 

\bsp	% typesetting comment
\label{lastpage}
\end{document}